\documentclass{emulateapj}

\usepackage{graphicx}
\usepackage{amssymb}

\slugcomment{Accepted for publication in ApJ}

\shorttitle{Star Formation in NGC 3521 \& M51a}
\shortauthors{Liu et al.}

\begin{document}


\title{The Super-linear Slope of the Spatially-Resolved Star Formation Law in NGC 3521 and NGC 5194 (M51a)}


\author{Guilin Liu\altaffilmark{1}}

\author{Jin Koda\altaffilmark{2}}

\author{Daniela Calzetti\altaffilmark{1}}

\author{Masayuki Fukuhara\altaffilmark{3}}


\author{Rieko Momose\altaffilmark{3,4}}

\altaffiltext{1}{Astronomy Department, University of Massachusetts, Amherst, MA 01003-9305, USA}
\altaffiltext{2}{Department of Physics and Astronomy, SUNY Stony Brook, Stony Brook, NY 11794-3800, USA}
\altaffiltext{3}{Department of Astronomy, University of Tokyo, Hongo, Bunkyo-ku, Tokyo 113-0033, Japan}
\altaffiltext{4}{National Astronomical Observatory of Japan, Mitaka, Tokyo 181-8588, Japan}

\begin{abstract}

We have conducted interferometric observations with the Combined Array for Research in Millimeter Astronomy 
(CARMA) and an on-the-fly mapping with the 45-m telescope at Nobeyama Radio Observatory (NRO45) in the 
CO ($J$=1--0) emission line of the nearby spiral galaxy NGC 3521. Using the new combined CARMA + NRO45 data 
of NGC 3521, together with similar data for NGC 5194 (M51a) and archival SINGS H$\alpha$, 24~$\mu$m, 
THINGS H {\sc i} and GALEX FUV data for these two galaxies, we investigate the empirical scaling law that 
connects the surface density of star formation rate (SFR) and cold gas (known as the Schmidt-Kennicutt law 
or S-K law) on a spatially-resolved basis, and find a super-linear slope for the S-K law when carefully 
subtracting the background emissions in the SFR image. We argue that plausibly deriving 
SFR maps of nearby galaxies requires the diffuse stellar and dust background emission to be subtracted carefully 
(especially in the mid-infrared and to a lesser extent in the FUV). 
Applying this approach, we perform a pixel-by-pixel analysis on both galaxies and quantitatively show that the 
controversial results whether the molecular S-K law (expressed as 
$\Sigma_{\rm SFR}\propto\Sigma_{\rm H_2}^{\gamma_{\rm H_2}}$) is super-linear or basically linear is a result of 
removing or preserving the local background. In both galaxies, the power index of the molecular S-K law is 
super-linear ($\gamma_{\rm H_2}\gtrsim1.5$) at the highest available resolution ($\sim$230 pc), and decreases 
monotonically for decreasing resolution. We also find in both galaxies that the scatter 
of the molecular S-K law ($\sigma_{\rm H_2}$) monotonically increases as the resolution becomes higher, indicating 
a trend for which the S-K law breaks down below some scale. Both $\gamma_{\rm H_2}$ and $\sigma_{\rm H_2}$ are 
systematically larger in M51a than in NGC 3521, but when plotted against the {\it de-projected} scale 
($\delta_{\rm dp}$), both quantities become highly consistent for the two galaxies, tentatively suggesting 
that the sub-kpc molecular S-K law in spiral galaxies depends only on the scale being considered, without 
varying amongst spiral galaxies. A logarithmic function $\gamma_{\rm H_2}=-1.1~\log[\delta_{\rm dp}/{\rm kpc}]+1.4$ 
and a linear relation $\sigma_{\rm H_2}=-0.2~[\delta_{\rm dp}/{\rm kpc}]+0.7$ are obtained through fitting to 
the M51a data, which describes both galaxies impressively well on sub-kpc scales. A larger sample of 
galaxies with better sensitivity, resolution and broader field of view are required to test the general 
applicability of these relations.


\end{abstract}





\section{INTRODUCTION}

Much of the information we obtain on galaxies and their evolution is the
consequence of the formation of stars, the process that depletes galaxies of their 
gas, dynamically energizes the stellar disks and enriches the interstellar medium 
with heavy elements through generations of evolving stars. Despite its fundamental 
importance on nearly all scales of astronomy, star formation is an exceedingly 
complex process to explore in the {\it ab initio} manner as a result of
the many entangling aspects including the dynamical, geometrical, magnetic and chemical 
properties of ISM, as well as the feedback that the young stellar objects exert on 
their surrounding environment by radiation, stellar wind and explosion of supernovae.

In spite of this difficulty, the well-established empirical scaling laws of star formation 
that connect the large-scale star formation rate (SFR) to interstellar medium (ISM) properties
already play a crucial role in establishing plausible scenarios for galaxy evolution, 
and serve as essential prescription in modeling and simulating galaxy formation and evolution.
Among them, the most widely adopted power-law relationship known as Schmidt law \citep{Schmidt59} 
links the volume or surface density of SFR to that of the gas \citep{Kennicutt89}. From an observational 
point of view, a variety of SFR indicators are currently used, along with gas density information derived 
from emission lines of molecules (e.g. CO, HCN), to trace the relation between the birthrate of stars and 
their environment. By averaging over the whole disk of individual galaxies and taking the total 
(atomic + molecular) cold gas into account, a single power-law scaling between the surface density of star 
formation and gas \citep[the Schmidt-Kennicutt law, abbreviated as the S-K law;][]{Kennicutt98} 
extending over five orders of magnitude in the gas density has been well established, taking the form 
$\rm \Sigma_{SFR} \propto \Sigma_{gas}^{1.4\pm0.15}$ where $\Sigma_{\rm SFR}$ (in 
$\rm M_{\odot}~yr^{-1}~kpc^{-2}$) and $\rm \Sigma_{\rm gas}$ (in $\rm M_{\odot}~pc^{-2}$) denote the 
disk-averaged SFR and gas surface density, respectively. 

Both parameters in this equation are, however, 
{\it global} quantities averaging over local values that may vary by several orders of magnitude. 
Further understanding of star formation on a more physical basis requires
scrutinizing galaxies at the sub-galactic scale. Limited by development of the relevant instrumentation, 
spatially-resolved investigations of the gas content of individual star-forming sites 
have become a front-line topic only in recent years \citep[e.g.,][]{Wong02, Boissier03, Heyer04, Kennicutt07, 
Bigiel08, Blanc09}. Spatially-resolved studies promise to yield insights into the physical processes driving 
the scaling laws, much more directly than the disk-averaged results. Measuring the star formation and gas at 
sub-kpc scales requires high resolution and sensitivity, which are currently available 
only for small or limited samples both for atomic hydrogen \citep[e.g., through THINGS;][]{Walter08} 
or molecular hydrogen \citep[via CO, e.g., HERACLES;][]{Leroy09}. Observational studies on the slope of 
the local S-K law in a handful of nearby galaxies result in power-law indices of 1--3 \citep[reviewed by,
e.g.,][]{Elmegreen02, Bigiel08}, where the cold gas mass in both atomic and molecular phases are 
taken into account. However, in recent times, it has become clearer that star formation activity,
at least on local scales, is much less correlated with H {\i} than with CO \citep[][hereafter K07 and B08, 
respectively]{Kennicutt07, Bigiel08}, which suggests that further investigations on the 
power law index of the S-K law should concentrate on the scaling relationship between SFR and molecular gas.

The availability of high resolution and sensitivity maps for CO as the most commonly used molecular hydrogen 
tracer thus provide the main actual limitation to sub-kpc studies of star formation scaling laws. 
In general, interferometric CO maps have higher resolution (typically a few arcseconds) and 
are powerful in resolving fine structures, but their limited field of view (FoV) and the shortcoming 
of missing the zero spacing flux often downgrade their scientific value; on the other hand, despite 
that single-dish observations are time-saving (especially if On-The-Fly mapping techniques are introduced) 
and cover much wider FoVs, their larger beams often smear out the details of star forming sites. 
In recent years, along with use of mosaicing that improves the FoV of interferometry,
the technique of compensating the absent short baselines components with single-dish data 
(at the price of losing some resolution) has been developed. With this technique mature, 
single-dish-compensated interferometric data of CO transition line maps become the best
available molecular gas tracer for investigations of star formation scaling laws. In this
direction, a new CARMA (the Combined Array for Research in Millimeter-wave Astronomy) 
\& NRO45 (Nobeyama Radio Observatory 45m telescope) CO survey of nearby galaxies is 
underway \citep{Koda09a} with the goal of better understanding the evolution of the ISM, galactic 
dynamics, and the star formation laws. Included in the sample of this CO survey, one of
our program galaxies, NGC 3521, is mapped by CARMA with a 3.65\arcsec$\times$3.14\arcsec~synthesized 
beam and by NRO45 using a 18.4\arcsec~beam, leading to a final resolution of 
5.43\arcsec$\times$4.68\arcsec~when the two sets of data are combined in {\it uv}-space with the 
new technique established by \citet{Koda09b}. This CARMA + NRO45 survey not only isolates typical 
GMCs with its $\sim$220 pc resolution at a distance of $\sim$8 Mpc \citep[a typical GMC separation 
is $\sim$200 pc in the Milky Way,][]{Koda06}, but also detects the interarm extended emission at 
a high sensitivity (1$\rm \sigma\sim3~M_{\odot}~pc^{-2}$).

The other crucial parameter in the resolved S-K law study is the SFR. Over the past twenty to thirty 
years, extensive efforts have been made to plausibly derive SFRs (see \citet{Calzetti10} for a review 
of the last decade and \citet{Kennicutt98} for earlier times). Conventional optical SFR tracers like 
H$\alpha$ often severely suffer from dust extinction 
\citep[$A_V\sim2.2$ mag in typical extragalactic H {\sc ii} regions,][]{Calzetti07}, 
which changes drastically from location to location and can be completely 
obscured in dense star forming regions where $A_V$ can reach $\sim$6 mag 
\citep[][for the case of M51a]{Scoville01}. Recent space-borne facilities have opened infrared (IR) 
and ultraviolet (UV) windows for star formation study. The Spitzer Infrared Nearby Galaxies Survey 
\citep[SINGS,][]{Kennicutt03} and the GALEX Nearby Galaxies Survey \citep[NGS,][]{Bianchi03a, Bianchi03b} 
enables us to image the details of dust obscured and unobscured star formation sites in parallel, 
so that both the dust reprocessed and directly emerging radiation from young stellar associations are measured. 
K07 and \citet{Calzetti07} have justified the feasibility of correcting the number of ionizing photons (as traced 
by the H$\alpha$ recombination line) for the effects of dust extinction by adding a weighted component 
from the Spitzer MIPS 24 $\mu$m luminosity in individual star forming regions. This H$\alpha$+24 $\mu$m 
SFR motivated \citet{Leroy08} and B08 to propose another composite SFR tracer that corrects dust attenuation 
of far-UV (FUV) surface brightness using the same mid-IR component (but with different weights). 

The extinction-corrected H$\alpha$ line emission has long been proved a reliable estimator of the 
rate of formation of young, massive stars, since stellar synthesis models show that for co-eval 
stellar populations with standard Initial Mass Functions that rate remains fairly constant for the 
first few million years ($\sim$5~Myr), and then abruptly fades away as massive stars become supernovae
\citep{Schaerer97}. 

In marked contrast, a stellar population can emit in the continuum UV (longward of 1300~\AA) for up 
to one order of magnitude longer timescale. The Local Volume Legacy (LVL) Survey team has found many 
dwarf galaxies that show little or no sign of dust and sporadic highly concentrated H$\alpha$ emission, 
yet show strong and extended UV emission and have red UV colors, consistent with stellar populations
at the age of a few 100 Myrs \citep{Dale09}. This complicates the comparison between H$\alpha$--based 
and UV-based SFR indicators. An in-depth discussion on the consistency and discrepancy of deriving SFR 
via H$\alpha$ and UV is presented in \citet{Lee09}. 

The issue of a molecular-only S-K law is only now starting to be tackled. Several earlier attempts were 
made by utilizing azimuthally averaged data. Employing this strategy, \citet{Wong02} studied seven 
CO-bright spiral galaxies and concluded that SFR is in direct proportion to molecular cloud density; 
\citet{Boissier03} compiled the data for 16 galaxies and found the slope $\sim$0.6--1.3 (though most 
data points seemingly favor a super-linear slope; see Figure 10 therein); \citet{Heyer04} examined 
the low-luminosity, molecule-poor M33 galaxy by combining 60 and 100 $\mu$m IR images with 
CO ($J$=1--0) single-dish line map, and suggested a power index of 1.36$\pm$0.08. 
More recently, SFR and molecular gas surface density have been compared on individual sub-kpc regions. 
K07 performed an analysis of star formation complexes in M51a (NGC 5194) and obtained 
a slope of 1.37$\pm$0.03 for the molecular S-K law at $\sim$500 pc resolution. On the other hand, 
B08 performed a pixel-by-pixel analysis (750 pc scale) for seven spiral galaxies and
found a power index 1.0$\pm$0.2 or a linear molecular S-K law. Thus it can be seen that the results
from different groups as yet fall into two categories: linear or super-linear scaling relation, which
is still now under debate. \citet{Wong02} would have obtained results similar to \citet{Heyer04} if 
radially dependent extinction corrections were applied to their H$\alpha$ data. 
B08 qualitatively suggests that the inconsistency between their results and K07 is 
likely an effect of removal of the local background emission in the K07 photometry of the SFR tracers. 
Clearly, background subtractions in the SFR maps is a crucial component for the investigation of star 
formation laws.

In parallel to these observational efforts, various star formation models have been constructed, 
generally assuming one mechanism that dominates the formation process. Among these models, a 
postulated constant star formation efficiency independent of the environment will lead to a linear 
relation \citep{Leroy08}, in which case the mere presence of the gas is sufficient for star formation;
a slope of 1.5 will be found if star formation is driven by large scale gravitational instabilities,
and thus occurs in a dynamical (free-fall) timescale of molecular gas \citep{Elmegreen02}; cloud-cloud 
collisions will drive the slope to be as steep as $\sim$2 \citep{Tasker09}; other considered mechanisms
include global resonances \citep{Wyse89}, galactic shear \citep{Hunter98}, dust shielding \citep{Gnedin09},
turbulence \citep{Krumholz09} and others \citep[for reviews, see][]{Elmegreen02, McKee07, Tan10}.

In this paper, we present a study of the sub-kpc S-K law in two nearby galaxies, NGC 3521 and 
M51a (NGC 5194). K07 and B08, despite sharing data sets for M51a, obtain different power law indices 
(super-linear vs. linear) of the molecular-only and total-hydrogen S-K laws. 
By employing a new method of local background subtraction, we show quantitatively that including 
or removing the local background in the maps tracing the SFR has a profound effect on the final 
results in M51a, as discussed in B08. By analyzing our high quality data, we infer that the linear 
correlation found by B08 for the molecular S-K law is a consequence of inclusion of stellar local 
background unrelated to current star formation. Throughout this paper, we base our analysis on the 
H$\alpha$+24 $\mu$m approach of calibrating SFRs, albeit the FUV+$24\mu$m SFR is included for the 
purpose of comparison. 

Among the extensive measurements of the distance to these two nearby galaxies, we adopt 
the values from the compilation by the LVL survey group: 8.03 Mpc for NGC 3521 \citep[from the flow-field 
corrected recessional velocity, assuming $H_0$=75~km~s$^{-1}$~Mpc$^{-1}$;][]{Kennicutt08}, and 8.00 Mpc 
for M51a \citep[by direct distance measurement of the M51 group,][]{Karachentsev04}.

NGC 3521 and M51a possess similar properties in several aspects. First of all, the disk-average 
SFR surface density is $\sim$0.0026 M$_{\odot}$ yr$^{-1}$ kpc$^{-2}$ and $\sim$0.0057 M$_{\odot}$ 
yr$^{-1}$ kpc$^{-2}$ for these two spirals, respectively \citep{Calzetti10}, characterizing
both as quiescently star-forming systems. Moreover, both galaxies are shown to be metal-rich 
from spectroscopic measurements, with 12+log(O/H) values 9.01 for NGC 3521 and 9.18 for M51a 
\citep{Moustakas10}. In addition, classified as SABbc in the Hubble sequence \citep[RC3,][]{RC3}, 
NGC 3521, with an inclination of 72.7$^{\circ}$ \citep{deBlok08}, is a flocculent-type spiral galaxy that 
has a tightly wound two-arm pattern \citep{Thornley96}, while its much less inclined peer M51a 
\citep[$i=42^{\circ}$,][]{Tamburro08}, a grand-design spiral of morphological type SAbc, shows a prominent 
two-arm configuration. The virtually identical size of their galactic disks is a consequence of having almost 
identical distances ($\sim$8 Mpc) and apparent size in the H$\alpha$ images that depicts unambiguously the 
spiral arms (284\arcsec~and 296\arcsec~in radius, respectively). These facts, when summed together, suggest 
that the two star--forming disks are very similar, with the major difference being their inclinations. 
This is an important ingredient for understanding the influence of spatial resolution on the resultant 
shape of star formation laws (see the result and discussion section).

The paper is organized as follows: section 2 describes the data and, especially for the CARMA/NRO45
data, provides some detail on the observations and data reduction; section 3 presents our new method
of automated local background removal which enables our pixel-by-pixel analysis to reproduce both
the results of K07 and B08 accurately, and discusses the necessity of local background removal; section 
4 is devoted to applying our strategy to both spirals to test the behavior of the sub-kpc star formation
laws; the results are discussed in section 5 and summarized in section 6.

\section{DATA}

\begin{figure*}
\plotone{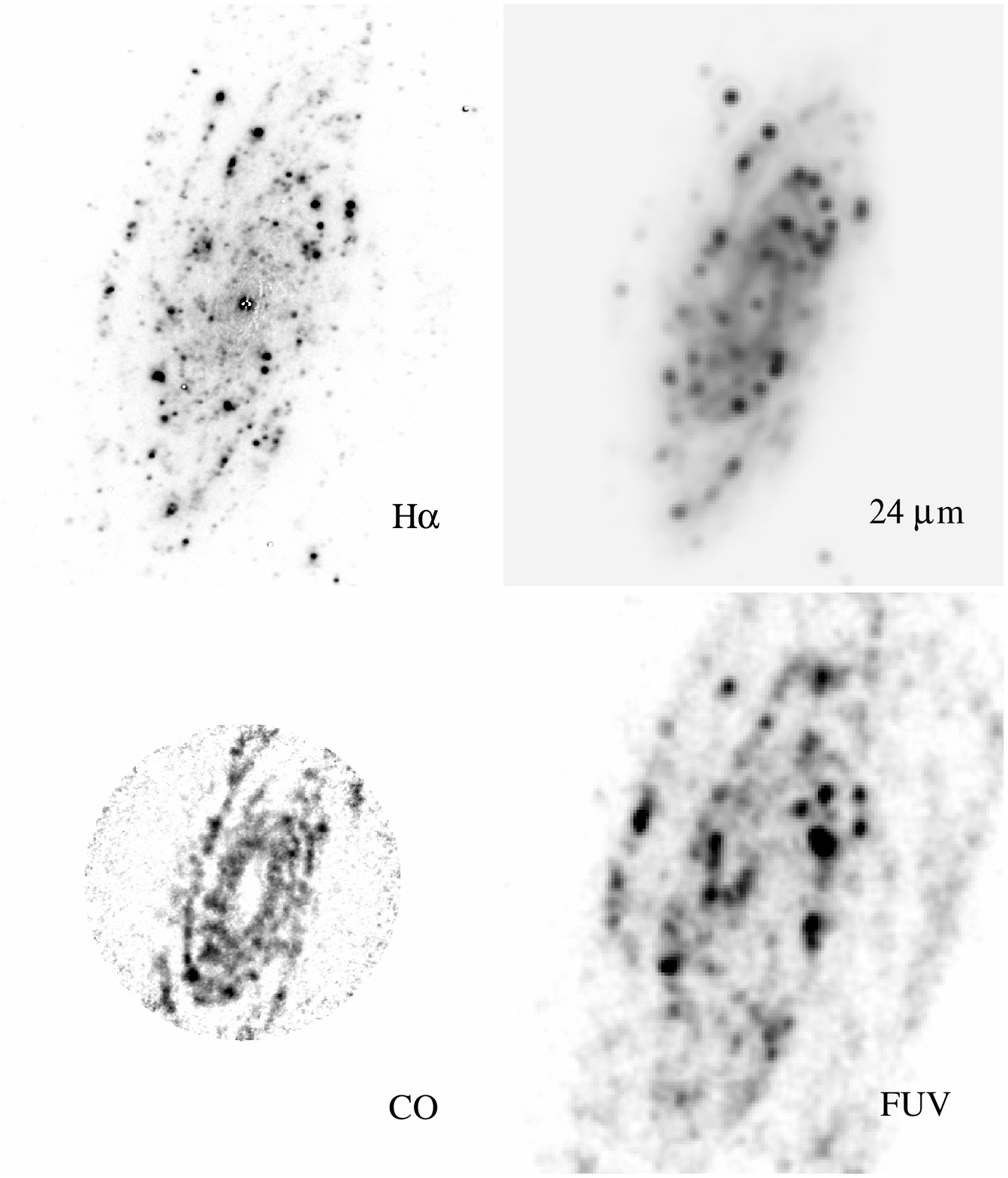}
\caption{The KPNO 2.1-m H$\alpha$, Spitzer MIPS 24 $\mu$m, CARMA CO ($J$=1--0) and GALEX FUV images of NGC 3521. 
North is up, East is left. The scale of each panel is $4.0\arcmin\times4.6\arcmin$, corresponding to 
9.3$\times$10.7 kpc$^2$.
\label{fig:images}}
\end{figure*}

In Figure~\ref{fig:images} we show the H$\alpha$ (KPNO 2.1-m telescope), 24 $\mu$m (Spitzer SINGS survey), 
CO (our CARMA observation) and FUV (GALEX NGS survey) images for NGC 3521. More technical
details of these data and the CARMA + NRO45 CO map of M51a are given in the following subsections. 
The H$\alpha$, 24 $\mu$m, BIMA SONG CO and FUV images of M51a, shared with K07 and B08,
have been thouroughly described in those papers and details will only be provided when necessary.
To compare our analysis with those in previous studies, we use the BIMA SONG data which were used in those 
studies, but our CARMA + NRO45 data are employed for further investigations.

\subsection{CARMA and NRO45 CO ($J$=1--0) Data} 

The CO ($J$=1--0) interferometric observations of NGC 3521 were undertaken using the Combined Array 
for Research in Millimeter Astronomy (CARMA) from February to March in 2009. CARMA is the combination
of the six 10 m antennas of the Owens Valley Radio Observatory (OVRO) millimeter interferometer 
and the nine 6 m antennas of the Berkeley-Illinois-Maryland Association (BIMA) interferometer. 
The entire optical disk of NGC 3521 (11.0$^{\prime}$$\times$5.1$^{\prime}$) was mosaiced in 19 
pointings with Nyquist sampling of the 10 m antenna beam (FWHM of 1$^{\prime}$ for the 115 
GHz CO $J$=1--0 line). The data were reduced and calibrated using the Multichannel Image 
Reconstruction, Image Analysis, and Display (MIRIAD) software package \citep{Sault95}.

We also obtained total power and short-spacing data with the 25-Beam Array Receiver System 
(BEARS) on the Nobeyama Radio Observatory 45 m telescope (NRO45, FWHM = 15\arcsec). 
Using the On-The-Fly observing mode \citep{Sawada08}, the data were oversampled on a 
5\arcsec~lattice and then re-gridded with a spheroidal smoothing function, 
resulting in a final resolution of 18.4\arcsec. Data reduction was performed using 
the NOSTAR data reduction package developed at the Nobeyama observatory. We constructed 
visibilities by deconvolving the NRO45 maps with the beam function (i.e., a convolution of 
the 15\arcsec~Gaussian and spheroidal function), and Fourier transforming them to 
the {\it uv}-space. After that, we combined the CARMA and NRO45 data in Fourier space, inverted 
the {\it uv} data using theoretical noise and natural weighting, and CLEANed the maps. 

Throughout this paper, we adopt a CO-to-H$_2$ conversion factor of 
$X_{\rm CO}$=2.8$\times$10$^{20}$ cm$^{-2}$ (K km s$^{-1}$)$^{-1}$ \citep{Bloemen86} for NGC 3521 
and M51a. The validity of this Galactic factor is justified by the super-solar abundance averaged 
over the galactic disks \citep[$\rm 12+log(O/H)=9.01\pm0.02$ for NGC 3521, and 9.18$\pm$0.01 for 
M51a,][]{Moustakas10}. Additionally, the small metallicity gradients across the disk justifies the 
application of a single value \citep{Bresolin04}.


For NGC 3521 at the distance of 8.03 Mpc, the angular resolution of the combined data 
(5.4\arcsec$\times$4.7\arcsec) corresponds to a physical scale of $\sim$220 pc, and the 
r.m.s. sensitivity (32 mJy beam$^{-1}$) in the 5.08 km s$^{-1}$ wide channels translates to 
a 3-$\sigma$ level of $\rm 2.4~M_{\odot}~pc^{-2}$ at this resolution (assuming $i$=72.7$^{\circ}$). 

\begin{figure*}
\includegraphics[scale=.6]{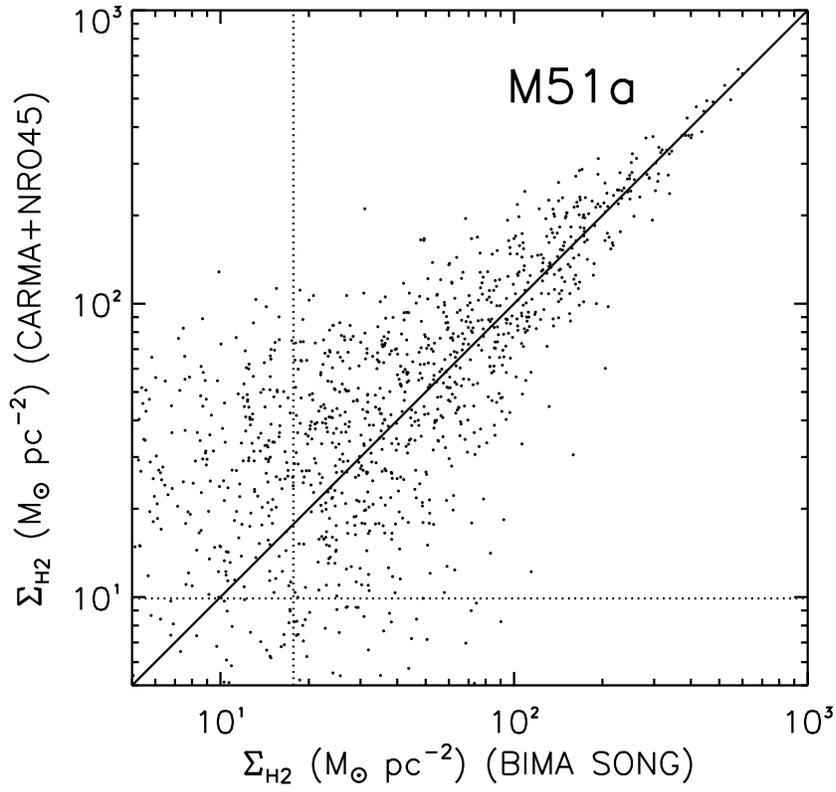}
\caption{Comparison of calculated molecular hydrogen surface densities in M51a between our 
CARMA + NRO45 and BIMA SONG CO measurements at the 230 pc (5.8\arcsec) resolution. The 3-$\sigma$ 
sensitivity limits of the SONG and our data at this resolution (18 and 9.9 $\rm M_{\odot}~pc^{-2}$,
respectively) are indicated by the dotted lines. The solid line indicates the expected 
unit slope in the case of perfect consistency. It is seen that the new CARMA + NRO45 data recover 
more CO flux at the faint end, and is deeper by a factor of $\sim$2 than the BIMA SONG 
data.
\label{fig:co_comp}}
\end{figure*}

As for M51a, the detailed information of the CARMA + NRO45 observations and data reduction is 
described in \citet{Koda09b} and \citet{Koda11}. At the final resolution of the combined map
(3.68\arcsec$\times$2.87\arcsec, corresponding to $\sim$140 pc at a distance of 8.00 Mpc), the r.m.s. 
sensitivity (40 mJy beam$^{-1}$) in the 5.08 km s$^{-1}$ wide channels translates to a 3-$\sigma$ level 
of $\rm 18~M_{\odot}~pc^{-2}$ (assuming $i$=42$^{\circ}$). For comparison, the typical size of a GMC in the Galaxy 
is 50 to several hundred parsecs \citep{Blitz93}. We show a comparison between our combined CARMA + NRO45 
and the BIMA SONG maps at a matched resolution, 5.8\arcsec~(corresponding to 230 pc for M51a) in 
Figure~\ref{fig:co_comp}. This highlights both the higher depth of our image and the fact that we recover 
substantial faint-level emission, which is absent from the BIMA SONG image.

\subsection{H$\alpha$ Emission-Line Images}

NGC 3521 was imaged using the 2.1 m telescope at Kitt Peak National Observatory as part 
of the SINGS ancillary data program \citep{Kennicutt03}. 
The H$\alpha$ line is contaminated by the doublet [N {\sc ii}] $\lambda$6548, 6584.
A ratio of $\lambda$6548, 6584/H$\alpha$=0.55$\pm$0.03 is adopted here to correct the
data, following the integrated spectrophotometric survey of nearby star-forming galaxies
accomplished by \citet{Moustakas10}. Our measurements on the final 
emission-line-only H$\alpha$ image has a PSF with a 1.4\arcsec~FWHM. The sensitivity 
limit of the emission-line image is 2.5$\times$10$^{-17}$ erg s$^{-1}$ cm$^{-2}$ arcsec$^{-2}$.

\subsection{Spitzer MIPS Images}

Spitzer MIPS 24~$\mu$m maps are available for NGC 3521, through the high level 
data products of the SINGS Legacy project \citep[Spitzer Infrared Nearby Galaxy 
Survey,][]{Kennicutt03}. The SINGS observation strategy, data reduction procedures, 
and map sensitivity limits are described in \citet{Kennicutt03} and \citet{Dale05}. 
For MIPS 24~$\mu$m maps, the diffraction-limited angular resolution is 5.7\arcsec~and 
the 1-$\sigma$ sensitivity limit 1.1$\times$10$^{-6}$ Jy arcsec$^{-2}$.

\subsection{VLA H {\sc i} Data}

The atomic hydrogen in NGC 3521 was mapped by the H {\sc i} Nearby Galaxies Survey 
\citep[THINGS,][]{Walter08} using the NRAO Very Large Array (VLA). The HI maps for our 
study are obtained through the THINGS public data release website\footnote{http://www.mpia-hd.mpg.de/THINGS/Overview.html}.
The H {\sc i} map, as a result of robust weighting, has a resolution of 8.19\arcsec$\times$6.41\arcsec. 
In each of the 5.2 km s$^{-1}$ channels, the 1--$\sigma$ sensitivity is 0.47 mJy 
beam$^{-1}$, corresponding to a column density of 5.8$\times$10$^{19}$cm$^{-2}$ or 
0.5 M$_{\odot}$pc$^{-2}$ in the integrated map. The calibration error for the H {\sc i} 
data are estimated to be $\sim$10\%. Further technical details are given in \citet{Walter08}.

\section{Methodology of Spatially-Resolved S-K Law Studies}

As mentioned in the Introduction, K07 and B08 find different results for the power law indices of 
the molecular S--K Law in M51a. B08 qualitatively point out that the difference arises because K07 
subtract a local H$\alpha$ and 24 $\mu$m background for each aperture, but different photometric 
approaches (K07: aperture photometry vs. B08: pixel-by-pixel) may also play a role. 
Meanwhile, such a test involves a highly tricky issue of local background removal. In this section,
we present a new approach of local background determination. By using it, we quantitatively 
show that removing or preserving the local background is sufficient to account for the discrepancy 
between K07 and B08. From here onward, we will indicate the fits to the $\Sigma_{\rm SFR}-\Sigma_{\rm H_2}$
relation as
\begin{equation}
\frac{\Sigma_{\rm SFR}}{\rm M_{\odot}~yr^{-1}~kpc^{-2}}=
A~\left({\frac{\Sigma_{\rm H_2}}{\rm M_{\odot}~pc^{-2}}}\right) ^{\gamma_{\rm H_2}}
\end{equation}
with $\gamma_{\rm H_2}$ indicating the power law exponent of the molecular gas surface density.
The dispersion about the fitted relation will be indicated with $\sigma_{\rm H_2}$. The analogous relation
for the total hydrogen gas will be indicated as
\begin{equation}
\frac{\Sigma_{\rm SFR}}{\rm M_{\odot}~yr^{-1}~kpc^{-2}} \propto
\left({\frac{\Sigma_{\rm H}}{\rm M_{\odot}~pc^{-2}}}\right) ^{\gamma_{\rm H}}
\end{equation}
where $\Sigma_{\rm H}=\Sigma_{\rm HI}+\Sigma_{\rm H_2}$. We also compare $\Sigma_{\rm SFR(FUV+24\mu m)}$
with $\Sigma_{\rm SFR(H\alpha+24\mu m)}$ through the relation
\begin{equation}
\Sigma_{\rm SFR(FUV+24\mu m)} \propto (\Sigma_{\rm SFR(H\alpha+24\mu m)})^{\gamma_{\rm SFR}}.
\end{equation}

\subsection{Local Background Subtraction}

In the hydrogen line, FUV and mid-IR images used to trace SFR, besides the compact clumpy structures 
which trace of H {\sc ii} regions and young massive stars, there exists an underlying 
diffuse component of stellar/dust emission, which is conventionally referred to as ``local background'' 
with slight ambiguity. This convention continues in this paper, but we stress that the so-called 
``local background'' here, apart from the local variation of sky or instrumental background, refers
to the physical detection of the emission from stars or interstellar dust.

Pixel-by-pixel analysis introduces significant contamination from this 
local background especially in lower resolution images, in which case the low luminosity pixels can 
be more strongly affected than those at the bright end, 
effectively overestimating the SFRs at the faint end and thus flattening the slope of the scaling laws.

Conventional approaches to determine the local background levels in galaxies include performing 
statistics in an annular region, or fitting the galactic image with an assumed two-dimensional 
analytical function (e.g., polynomial, exponential, S\'{e}rsic profile or a combination of several 
galactic components). These methods work reasonably well in general but often fail in 
nearby galaxies because of the large dynamical range and crowding of clustered regions. 
\citet{Calzetti05} and K07 defined 12 rectangular regions in the M51a images that they used as their
``background'' areas. Since it is not sensitive to local variations on smaller scales, this ad-hoc strategy 
produces reasonable results (see next subsection). However, a more automated approach is expected to enhance 
the reproduceability and robustness of the results. \citet{Blanc09} identify and remove the diffuse ionized 
gas (DIG) from the H$\alpha$ image of M51a statistically, based on \citet{Madsen06} which demonstrate that the 
DIG in the Galaxy has a differenct [S II]/H$\alpha$ ratio (0.34) from that of Galactic H {\sc ii} regions (0.11). 
The \citet{Blanc09} procedure has the advantage of being astrophysically motivated, but the DIG fraction 
recovered by these authors (11\%) is smaller than that recovered by photometric measurements either 
when the whole galaxy \citep[40--50\%, e.g.,][]{Greenawalt98,Thilker00} or the central part of M51a 
is considered \citep[we find the fraction to be 32\% in the same central part covered by][using {\sl HIIphot}]{Blanc09}.

In this work, we distinguish the compact component (presumably H {\sc ii} regions) from diffuse stellar/dust 
emission through morphological examination. This strategy is inevitably arbitrary to some extent, but its 
basic motivation has a strong physical underpinning. The properties of the diffuse H$\alpha$, 
24 $\mu$m and UV emission have been extensively discussed by the existing literature, which is briefly 
reviewed in the discussion section. When such a strategy is being applied, the crowding of sources is the 
major obstacle to accurate background determination, but it can be circumvented by removing the compact 
clumps instead, which is done by adapting the algorithm of the IDL software {\sl HIIphot} developed by 
\citet{Thilker00} for our purposes in this work. The algorithm can be summarized as follows. On an H$\alpha$ 
image of a galaxy, it identifies H {\sc ii} regions by smoothing the image with kernels of different sizes, 
and utilizes object recognition techniques to identify significant peaks (``seeds''). After 
that, an iterative procedure is started allowing for area growth from these ``seeds''. The boundaries 
are built up at successively fainter isophotal levels that keep growing until either boundaries of other 
regions are encountered, or a pre-established lower limit to the gradient in surface brightness is reached 
implying its arrival at the background level. Two-dimension interpolations are then carried out by fitting 
the background pixels around each H {\sc ii} region to estimate the local background behind the source. 

Originally designed for automated photometric characterization and statistical analysis of H {\sc ii} regions 
and their luminosity functions in H$\alpha$ images, the algorithm of {\sl HIIphot} can be easily 
adapted to 24 $\mu$m and FUV images. We tested our implementation of {\sl HIIphot} on the HST H$\alpha$ image
of M51a, and produced an H {\sc ii} region luminosity function which is very similar to that of 
\citet{Scoville01} \citep{Liu11}.
It should be noticed that our purpose of extracting star-forming regions 
against the local background is somewhat different from conventional H$\alpha$ data analysis which aims at
removing any contamination from diffuse components in H {\sc ii} regions. The lower limit gradient in surface 
brightness for the iterative growing procedure from the ``seeds'' is thus chosen to be sufficiently low
(lower than what is typically chosen for H {\sc ii} region analysis) that we are confident to have reached
an appropriate background level, but still sufficiently above zero to prevent growing deep into ambient areas.

Because of the much lower contribution from the diffuse emission in H$\alpha$ than in mid-IR and FUV, the removal 
of the H$\alpha$ local background is done conservatively, aiming at a background level to be reached only 
in the ambient area between the outer spiral arms.  
For the purpose of testing this approach, we also use a constant global background determined using several 
external H {\sc ii} regions in deriving S-K laws, and the resultant slopes differ minimally 
($<0.1$, typically $\sim$0.05).

\begin{figure*}
    \includegraphics[totalheight=2.5cm,angle=0,origin=c,scale=2.4]{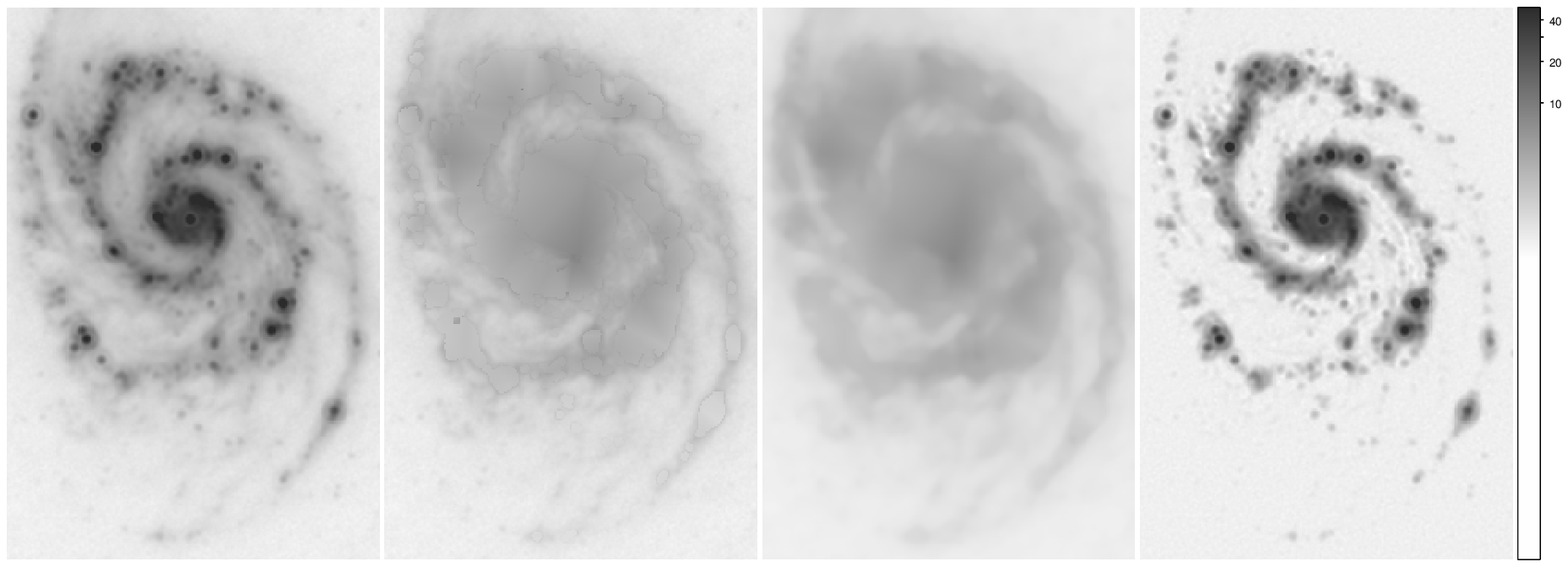}
    \includegraphics[totalheight=2.5cm,angle=0,origin=c,scale=2.4]{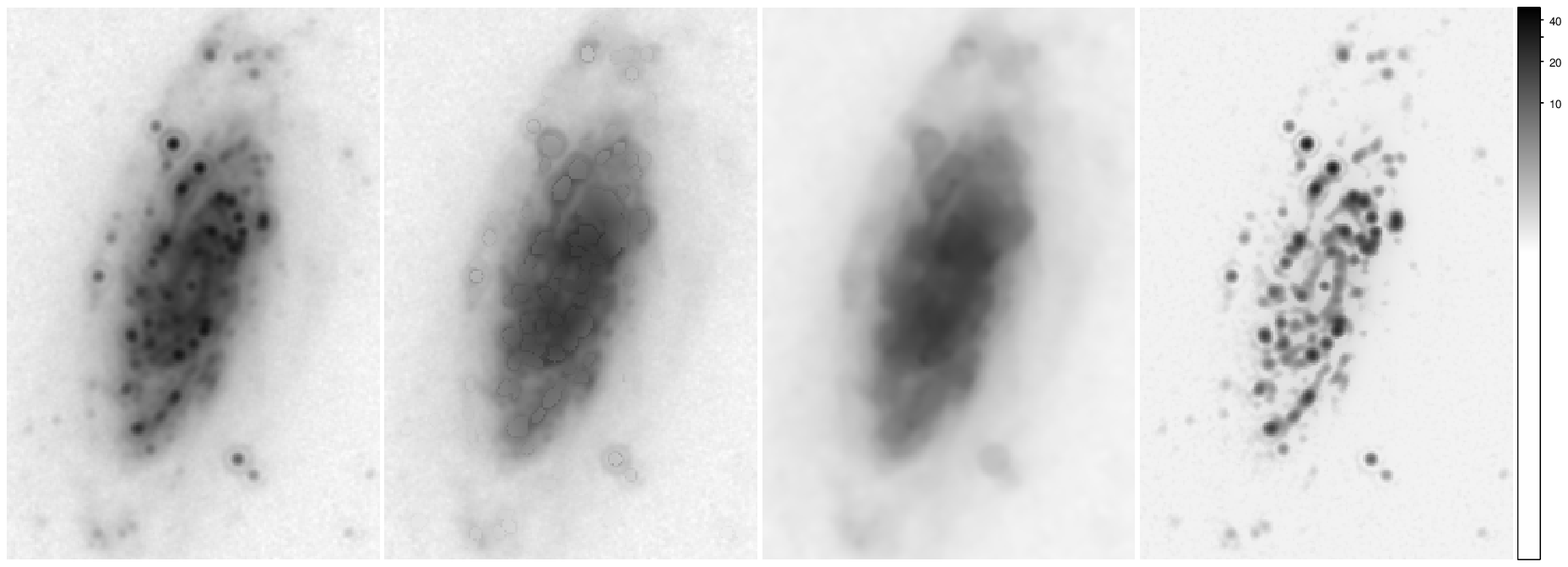}
    \caption{The products of our local background removal strategy applied on the
24 $\mu$m images of M51a (top row) and NGC 3521 (bottom row). In each row, the 
compact emission component of the original image (first column) is identified and
removed by adopting the algorithm of {\sl HIIphot}, then the resultant image (second 
column) is smoothed by replacing each pixel with the median of a moving box so
that no sharp features exist. The background image is thus created (third column)
and subtracted from the original image, so that the final background-free image 
(last column) is prepared for our analysis. All images for each galaxy share common
logarithmetic grey scales displaying surface brightness in units of MJy sr$^{-1}$.}
\label{fig:bkgd}
\end{figure*}

The established source-free image is not smooth, but contains sharp edges at the boundaries of
sources. We therefore smooth the image by calculating the median of the pixels 
enclosed in a moving box and replacing the central pixel with the median. Because of the robustness of
the median in statistics, this smoothing technique leads to a relatively stable result and suffers much 
less from the steep local variations which will affect an image smoothed by kernel convolution. 
The size of the box is chosen as a compromise between minimization of the crater-like artifacts left by
the removed H {\sc ii} regions and minimization of detail loss. The final resulting ``diffuse emission'' 
map is then ready for subtraction from the original image. The background-free images, along with the 
background images before and after the median smoothing, are shown in Figure~\ref{fig:bkgd} for the
purpose of comparison. As can be seen from the figure, the background-free image mostly contains
clustered sources, and the disk-like diffuse emission has been removed.

\subsection{Linear Relation without Local Background Subtraction}

B08 performed a pixel-by-pixel analysis on 7 spiral galaxies detected in CO ($J$=2--1) at a resolution of 
750 pc including M51a. The employed SFR tracer in their study is a combination of far-UV and 24 $\mu$m 
luminosities without local background subtraction and whose coefficients are adjusted to be consistent 
with the H$\alpha$+24 $\mu$m SFR tracer at 750 pc resolution, the applicability of which has been justified 
in K07 and \citet{Calzetti07}. No correlation between SFR and H {\sc i} surface density is seen, a similar 
result to K07. In contrary to the super-linear relationship found by K07 (power law index 
$\gamma_{\rm H_2}=1.37\pm0.03$), B08 find the surface density of SFRs to be proportional to that of molecular 
gas (power index $\gamma_{\rm H_2}=1.0\pm0.2$). We notice that B08 carried out 
their analysis using CO ($J$=2--1) images from The HERA CO Line Extragalactic Survey 
\citep[HERACLES;][]{Leroy09} except for M51a, for which they share the same data set as K07, including the 
CO ($J$=1--0) image from BIMA Survey of Nearby Galaxies \citep[BIMA SONG;][]{Helfer03}. This enables us to 
make a direct comparison of the two approaches of measurements adopted in these two studies to identify the 
reason for the discrepancy of the obtained power law indices. 

The KPNO H$\alpha$, SINGS 24 $\mu$m, GALEX FUV, THINGS 21-cm H {\sc i} and BIMA SONG CO ($J$=1--0) 
images of M51a have resolutions of 1.9\arcsec, 5.7\arcsec, 4.6\arcsec, 5.82\arcsec$\times$5.56\arcsec~and 5.8\arcsec$\times$5.1\arcsec, respectively. The 3-$\sigma$ sensitivity is 
1.3 M$_{\odot}$ pc$^{-2}$ for the H {\sc i} map (K07), and 13 M$_{\odot}$ pc$^{-2}$ for 
the CO map \citep{Helfer03}. The 1-$\sigma$ detection limits of the 24 $\mu$m (1.1$\times$10$^{-6}$ 
Jy arcsec$^{-2}$), H$\alpha$ (1.8$\times$10$^{-17}$ erg s$^{-1}$ cm$^{-2}$ arcsec$^{-2}$) and 
FUV images (3.6$\times$10$^{-19}$ erg s$^{-1}$ cm$^{-2}$\AA$^{-1}$ arcsec$^{-2}$) \citep{Calzetti05} 
lead to a corresponding SFR$_{\rm H\alpha+24\mu m}$ of 1.0$\times10^{-5}$ M$_{\odot}$~yr$^{-1}$~kpc$^{-2}$ 
and a SFR$_{\rm FUV+24\mu m}$ of $1.7\times10^{-5}$ M$_{\odot}$~yr$^{-1}$~kpc$^{-2}$ at 750 pc (19.3\arcsec) 
resolution. Here we have assumed an IMF given by \citet{Kroupa01} to derive the SFR surface density 
(in units of $\rm M_{\odot}~yr^{-1}~kpc^{-2}$) following \citet{Calzetti07} and \citet{Leroy08}:
\begin{equation}
 {\rm SFR}_{\rm H\alpha+24\mu m}=5.3 \times 10^{-42}~(L_{\rm H\alpha}+0.031~L_{\rm 24\mu m}),
\end{equation}
\begin{equation}
 {\rm SFR}_{\rm FUV + 24\mu m} = 3.4 \times 10^{-44}~(L_{\rm FUV} + 6.0~L_{\rm 24\mu m}),
\end{equation}
where $L_{\rm 24\mu m}$ and $L_{\rm FUV}$, both defined as $\lambda L_{\lambda}$ and 
in units of erg s$^{-1}$, are referenced to the central wavelength of the Spitzer 24 $\mu$m
($\lambda$=23.68$\mu$m) and the GALEX FUV ($\lambda$=1528\AA) filters, respectively. 

We first convolve these M51a images with circular or elliptical 2-d gaussian kernels so that all the 
images are downgraded to a common physical resolution of 750 pc, the working scale of 
B08, for the purpose of comparison. Aligned to the astrometric frame of the 24 $\mu$m 
image which has the coarsest grid (1.5\arcsec), they are then regrided to a pixel scale of 750 pc 
to avoid sub-PSF sampling. The data are then converted to physical 
quantities, correlated pixel-by-pixel, and analyzed by linear fitting. In this study, the linear
regressions are performed with the ordinary least square bisector method (same as K07 and B08), as 
recommended by \citet{Isobe90}.

\begin{figure*}
\plotone{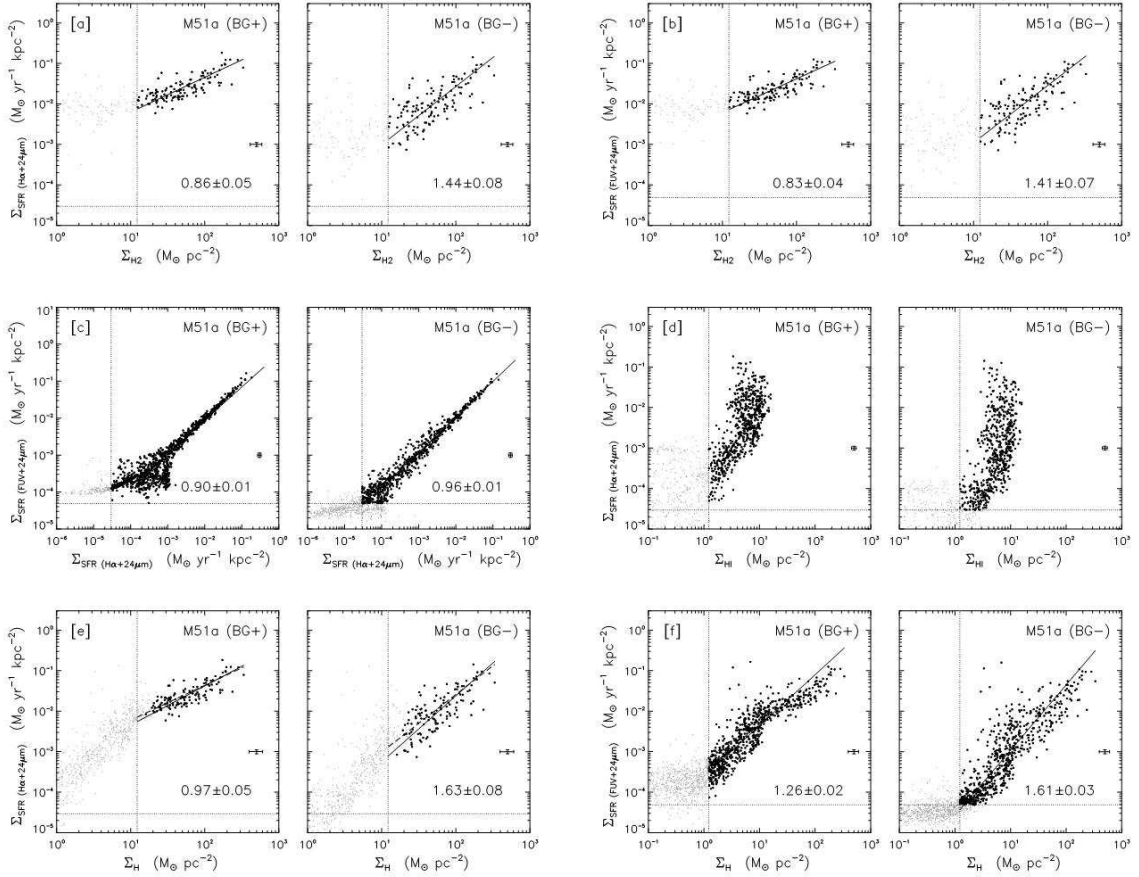}
    \caption{The correlations for M51a between derived physical quantities at 750 pc resolution
(at a distance of 8.00 Mpc). For the left panel in each pair the local backgrounds in the
H$\alpha$, 24 $\mu$m and FUV images are not removed (denoted by ``BG+''), but are removed in 
the right panel (``BG$-$''). The fitted slopes are indicated at the bottom-right of each panel.
[a] the molecular-only S-K law with SFR derived from H$\alpha$+24 $\mu$m; 
[b] the molecular-only S-K law with SFR derived from FUV+24 $\mu$m; [c] the correlation between SFRs 
derived from H$\alpha$+24 $\mu$m and FUV+24 $\mu$m, respectively; [d] the relation of H$\alpha$+24 $\mu$m 
SFR vs. H {\sc i} surface density; [e] the total hydrogen S-K law with SFR derived from H$\alpha$+24 $\mu$m; 
[f] the total hydrogen S-K law with SFR derived from FUV+24 $\mu$m. In each panel, the horizontal and 
vertical dotted lines locate the 3-$\sigma$ sensitivity limit of the two involved quantities. Note that
in general, the sensitivity limit of $\Sigma_{\rm SFR}$ is not the limiting factor for our fitted range,
while $\Sigma_{\rm gas}$ is. Those data points with sufficient S/N ratio for linear bisector fittings 
are shown as solid black dots, in contrast to the low S/N light grey dots. In panel [e], the 
solid line corresponds to the fit using points where CO, H {\sc i} and SFR are independently above 3$\sigma$ 
(best-fit power index $\gamma_{\rm H}=1.63\pm0.08$), while the dashed line results from requiring the total hydrogen surface 
density and SFR to be above 3$\sigma$ (best-fit $\gamma_{\rm H}=1.42\pm0.06$).}
\label{fig:m51_bima_comp}
\end{figure*}

The results for M51a are summarized in Figure~\ref{fig:m51_bima_comp} (the left of the panel pairs). 
At this resolution, we reproduce the results of B08 ($\gamma_{\rm H_2,B08}$=0.84) when 
$\rm FUV+24~\mu m$ SFRs are correlated to molecular gas surface densities ($\gamma_{\rm H_2}$=0.83$\pm$0.04, 
Figure~\ref{fig:m51_bima_comp}-b). we find a similar sub-linear correlation ($\gamma_{\rm H_2}$=0.86$\pm$0.05) 
when using H$\alpha$+24 $\mu$m as a SFR indicator for the molecular-only S-K law (Figure~\ref{fig:m51_bima_comp}-a),
because of the tight, close-to-linear correlation between SFR$_{\rm FUV+24\mu m}$ and SFR$_{\rm H\alpha+24\mu m}$ 
in the $\rm SFR\gtrsim10^{-3}~\rm M_{\odot}~yr^{-1}~kpc^{-2}$ regime 
\citep[a result of the calibration by][Figure~\ref{fig:m51_bima_comp}-c]{Leroy08}.
Nevertheless, the two SFR tracers correlate poorly with each other when SFR is lower than 
$10^{-3}~\rm M_{\odot}~yr^{-1}~kpc^{-2}$, hinting at a discrepancy of deriving H$\alpha$ and UV
at low surface brightnesses. All these linear fits have taken into account 
only those data points above their corresponding 3-$\sigma$ sensitivity limit which are shown 
in each panel as horizontal or vertical dotted lines. 

Little or no correlation is found between SFR$_{\rm FUV+24\mu m}$ or SFR$_{\rm H\alpha+24\mu m}$ with 
the atomic gas surface density, and we observe a saturation at $\rm \Sigma_{HI}\sim 10~M_{\odot}~pc^{-2}$. 
Both effects had been already observed by both K07 and B08. An evident ``kink'' is seen at the transition 
from atomic- to molecular-dominated regimes when both components are combined to make a 
SFR -- $\Sigma_{\rm H}$ comparison (Figure~\ref{fig:m51_bima_comp}-e and \ref{fig:m51_bima_comp}-f). 
Compared to Figure~\ref{fig:m51_bima_comp}-a or \ref{fig:m51_bima_comp}-b, the slope hardly changes in 
Figure~\ref{fig:m51_bima_comp}-e ($\gamma_{\rm H}$=0.86$\pm$0.03, dashed line) where we restrict the fitting 
to pixels in which the total hydrogen surface density and SFR are above 3$\sigma$. However, when 
extended to the outer disk by assuming little or no existing molecules, the linear fit gives 
$\gamma_{\rm H}$=1.26$\pm$0.02, steeper than $\gamma_{\rm H}$=1.11 found by B08, a consequence of the fact that
we do not mask the regions outside $D_{25}$. The results for M51a presented by B08 are therefore 
thouroughly reproduced by our independent pixel-by-pixel analysis. 

We perform a similar comparison to B08's result for NGC 3521, although we use our CARMA + NRO45 CO ($J$=1-0)
map, while those authors used their HERACLES CO ($J$=2-1) data. Since we adopt a smaller distance for this 
galaxy than B08 (8.03 Mpc vs. 10.7 Mpc for those authors), their assumed 750~pc pixels correspond in reality 
to 1~kpc pixels. We thus re-grid NGC 3521 to 1~kpc pixels and obtain a molecular-only power law index 
$\gamma_{\rm H_2}=0.90\pm0.13$ when SFR is derived from H$\alpha$ and 24 $\mu$m data, or $\gamma_{\rm H_2}=0.92\pm0.12$ 
when FUV is used instead (B08 find $\gamma_{\rm H_2,B08}=0.95$). The SFRs derived in these two different ways 
coincides with each other, SFR$_{\rm FUV+24\mu m} \propto$ SFR$_{\rm H\alpha+24\mu m}^{1.02 \pm 0.02}$.
The scaling relation for total hydrogen gas content gives $\gamma_{\rm H}=2.06\pm0.31$ (without masking 
any part of the H {\sc i} map), similar to B08's value of 2.12. We infer that the two different CO maps 
($J$=1-0 and $J$=2-1) do not lead to discrepancies in the resulting S-K law, at least at the spatial scales 
investigated for this galaxy. 

\subsection{Super-Linear Relation with Local Background Subtraction}

By employing both the extinction-corrected Pa$\alpha$ line image and the weighted sum 
of H$\alpha$ and 24 $\mu$m luminosities as an unbiased SFR tracer, K07 performed photometry 
using apertures with 13\arcsec~diameter ($=520$ pc) on 257 positions in M51a where star formation 
is detected. 
The authors identified 12 rectangular areas and fit the local background for subtraction. This strategy
was applied to both the hydrogen line image and the 24 $\mu$m map. As a result, a best-fit 
slope of $\gamma_{\rm H}=1.56\pm0.04$ was found when the specific SFR is correlated to the total gas surface
density on 520 pc scales; the slope decreases to 1.37$\pm$0.03 on 1850 pc scales. 
Additionally, virtually no correlation is seen between the local SFR surface density and the
H {\sc i} surface density which saturates at log [$\Sigma_{\rm HI}$/M$_{\odot}$~pc$^{-2}$]$\sim$1, 
while a strong correlation with a shallower best-fit slope of $\gamma_{\rm H_2}=1.37\pm0.03$ exists for 
the molecular S-K law on 520 pc.

The actual consistency of the results from K07 and B08 can be shown using the background-subtracted 
24 $\mu$m, FUV and H$\alpha$ images of M51a, combined with the BIMA SONG CO map. 
We switch our pixel-by-pixel analysis to 507 pc (adopting a distance of $D$=8.0 Mpc), the scale 
examined in K07 (520 pc, assuming $D$=8.2 Mpc), to retrieve their results, where 
OLS bisector fittings are performed again, with the local background subtracted in the SFR map. 
Our 500 pc molecular-only S-K law in M51a has a slope of $\gamma_{\rm H_2}=1.48\pm0.06$, slightly 
steeper than but only 1.3$\sigma$ away from what is obtained by K07 through 
$\rm H\alpha+24~\mu m$ SFR, $\gamma_{\rm H_2,K07}=1.37\pm0.03$. 
The best-fit power relation between the two SFR tracers is given by 
SFR$_{\rm FUV+24~\mu m} \propto$ SFR$_{\rm H\alpha+24\mu m}^{0.95\pm0.01}$. 
This calibration now has been improved relative to that from the unsubtracted images (section 3.2)
by a remarkably weakened scatter at the faint end, and the
tight correlation now extends down to $\rm SFR\sim10^{-4}~\rm M_{\odot}~yr^{-1}~kpc^{-2}$, implying
the applicability of FUV + 24$\mu$m as a SFR tracer as long as the local background is properly removed. 
Not surprisingly, $\gamma_{\rm H_2}$ changes insignificantly to 1.47$\pm$0.05 if unobscured SFRs are measured 
by substituting H$\alpha$ with FUV luminosity. A even stronger sign of no correlation between 
$\Sigma_{\rm HI}$ and $\Sigma_{\rm SFR}$ is seen. For the total gas, we find a best-fit 
$\gamma_{\rm H}$=1.52$\pm$0.05 when a 3-$\sigma$ sensitivity limit cutoff is adopted for the total 
hydrogen mass surface density, virtually identical to the K07 result. However, from a practical 
point of view, only those data points with both H {\sc i} and CO detections above their respective 
3$\sigma$ should be taken into account for the linear fit, just as K07 did in their study of total gas 
surface density scaling relation. Such a criteria results in a steeper slope $\gamma_{\rm H}$=1.66$\pm$0.06, 
still consistent with K07. Hence, all the results that K07 obtained by performing aperture photometry are
reproduced with our pixel-by-pixel analysis on the background subtracted images. 

In conclusion, using the same set of data as K07 and B08, our analysis accurately reproduces the results 
of both investigations. We hereby conclude that at least for the case of M51a, the crux of their discrepancy 
lies in the preservation/elimination of the local background in 24 $\mu$m, FUV maps and H$\alpha$ images, 
ruling out the different employed measuring strategies (aperture photometry vs. pixel-by-pixel analysis) 
and any other factors as possible important causes.

\section{Star Formation in M51a and NGC 3521 on Sub-kpc Scales}

\begin{figure*}
\plotone{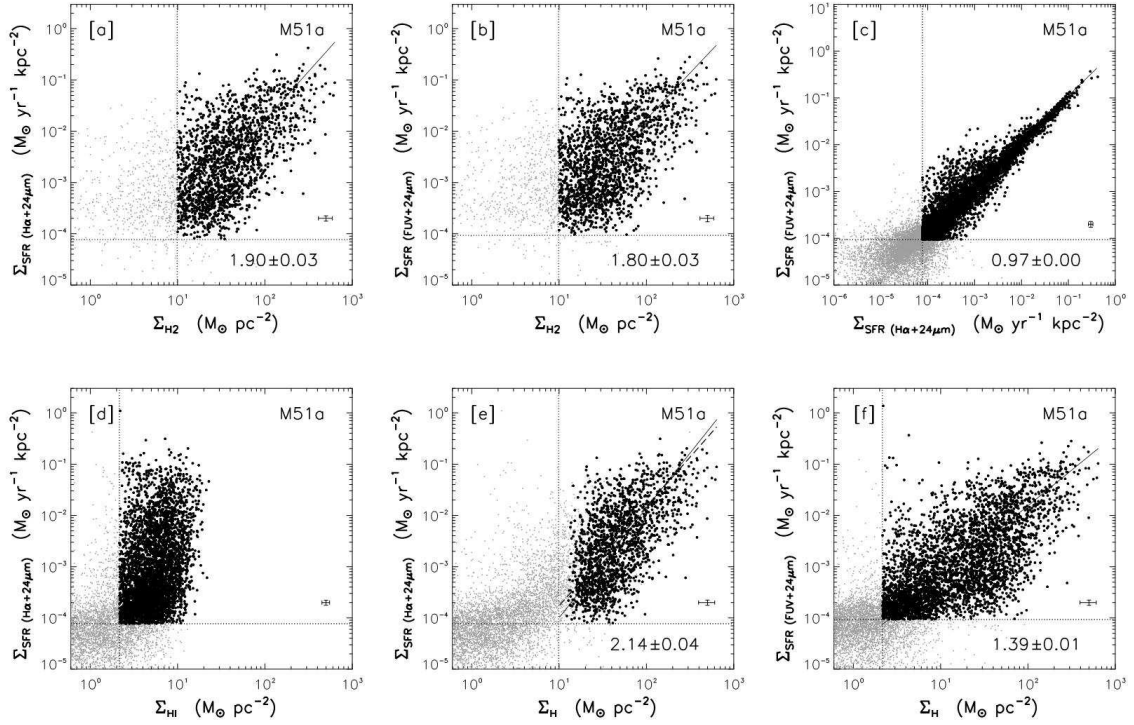}
    \caption{The scaling relations of star formation for M51a 
at the highest available resolution (230 pc). The local backgrounds in H$\alpha$, 24 $\mu$m 
and FUV images have been removed, following our strategy. The symbols, notes and involved 
physical quantities are identical to Figure~\ref{fig:m51_bima_comp}. 
}
\label{fig:m51_sk}
\end{figure*}

\begin{figure*}
\plotone{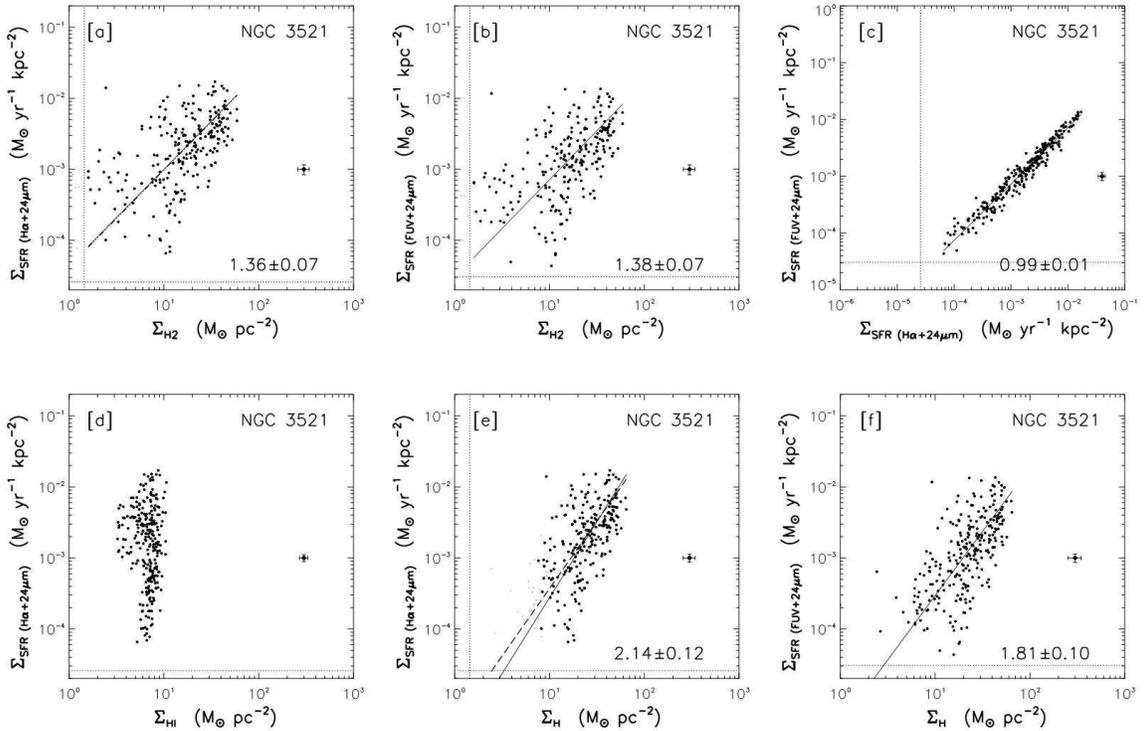}
    \caption{The scaling relations of star formation for NGC 3521 at the highest available 
resolution for H {\sc i} data (320 pc). The local backgrounds in H$\alpha$, 24 $\mu$m 
and FUV images have been removed, following our strategy. The symbols, notes and involved 
physical quantities are identical to Figure~\ref{fig:m51_bima_comp}. All quantities are compared 
inside the FoV of CARMA CO data, which is $\sim30\%$ of the optical area defined by $D_{25}$ used 
by B08.}
\label{fig:n3521_sk}
\end{figure*}

The BIMA SONG CO map of M51a has been used so far for the purpose of comparison only, and is now replaced 
with the more recent CARMA + NRO45 data to actually probe the spatially-resolved S-K law.
For NGC 3521, the spatial resolution of our data set is limited by that of the {\sc H i} map 
(8.19\arcsec$\times$6.41\arcsec), corresponding to a physical scale of 320~pc at the adopted distance, 
8.03 Mpc. However, the resolution limit of molecular-only S-K law can be pushed to smaller scale values, to
220 pc or 5.7\arcsec, the PSF FWHM of the 24 $\mu$m image. As for M51a, the similar resolutions of its H {\sc i} 
map (5.82\arcsec$\times$5.56\arcsec) and 24 $\mu$m image set the accessible limit to be $\sim230$ pc.
The {\sc H i} and 24~$\mu$m maps, which have the coarsest pixels (1.5\arcsec), define the grid onto which 
images are registered. We then smooth all the images to make the PSF circular with the same resolution
and rescale the pixels to match that resolution before converting flux and surface brightness units to physical 
quantities. An inclination angle of 72.7$^{\circ}$ (42$^{\circ}$) results in a correction factor of 0.29 (0.74) 
for projected surface densities in NGC 3521 (M51a). The CO data, having a lower sensitivity limit than {\sc H i} 
by a factor of $\sim$2.3, set our limiting sensitivity to the cold gas surface density measurement. Justified by 
the results of the previous sections, we now apply a pixel-by-pixel analysis to both galaxies, M51a and 
NGC 3521, utilizing SFR images where the diffuse emission/local background has been removed. The S-K law of 
M51a and NGC 3521 are shown in Figure~\ref{fig:m51_sk} and Figure~\ref{fig:n3521_sk} at their highest available 
resolution, respectively. 

\begin{figure*}
    \includegraphics[scale=.4,clip,trim=0cm 0cm 0cm 0cm]{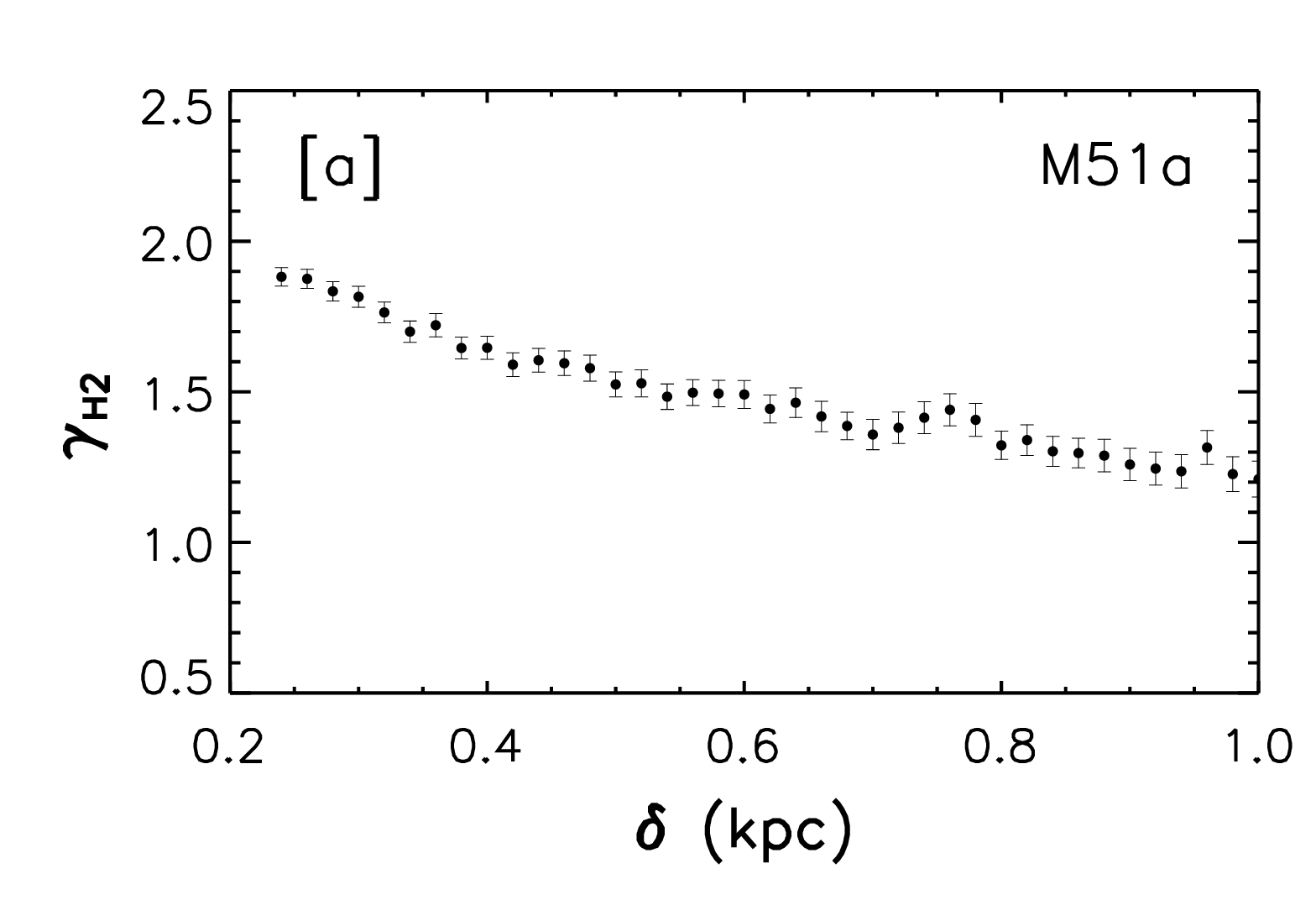}%
    \includegraphics[scale=.4,clip,trim=2.5cm 0cm 0cm 0cm]{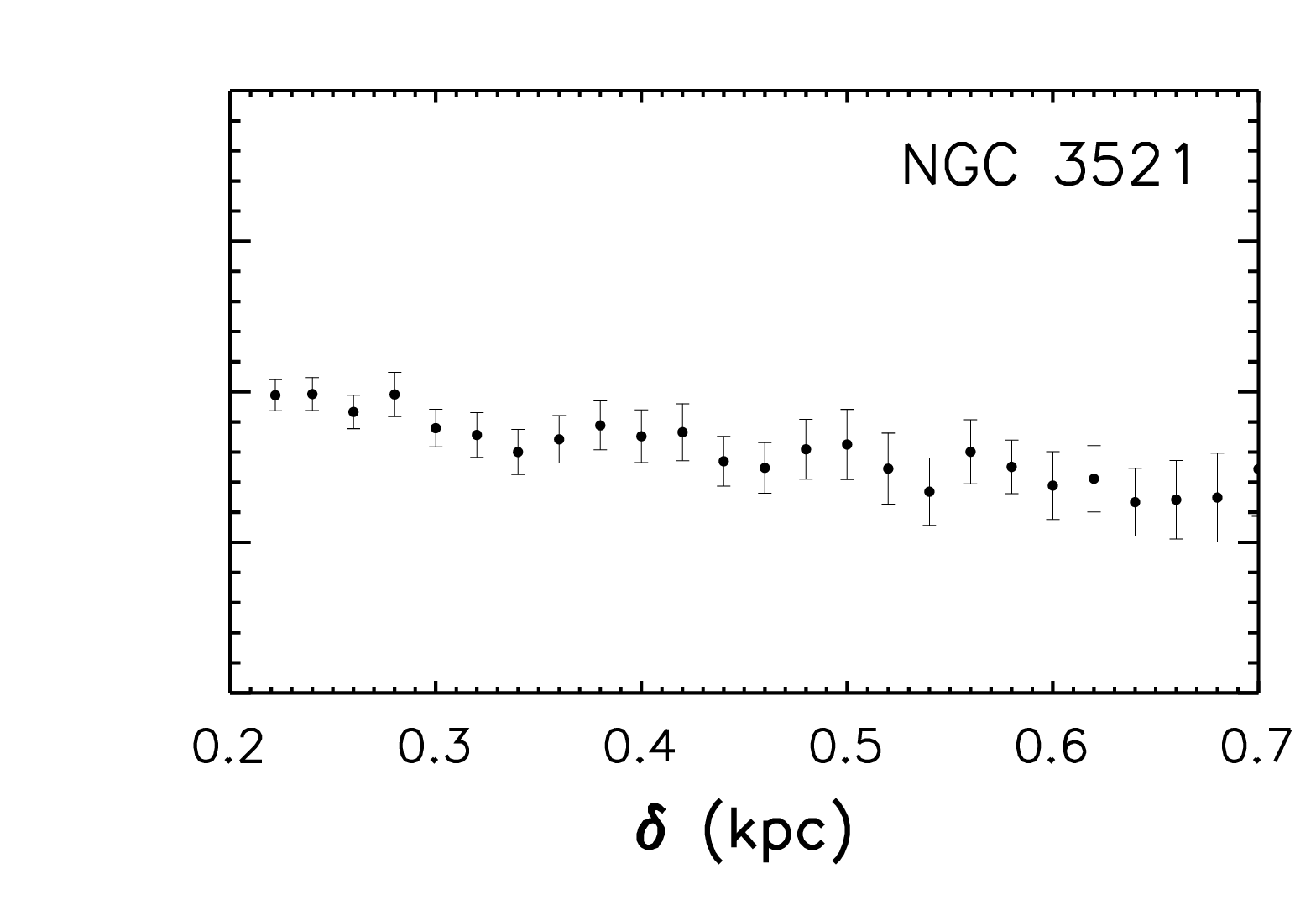}
    \includegraphics[scale=.4,clip,trim=0cm 0cm 0cm 0cm]{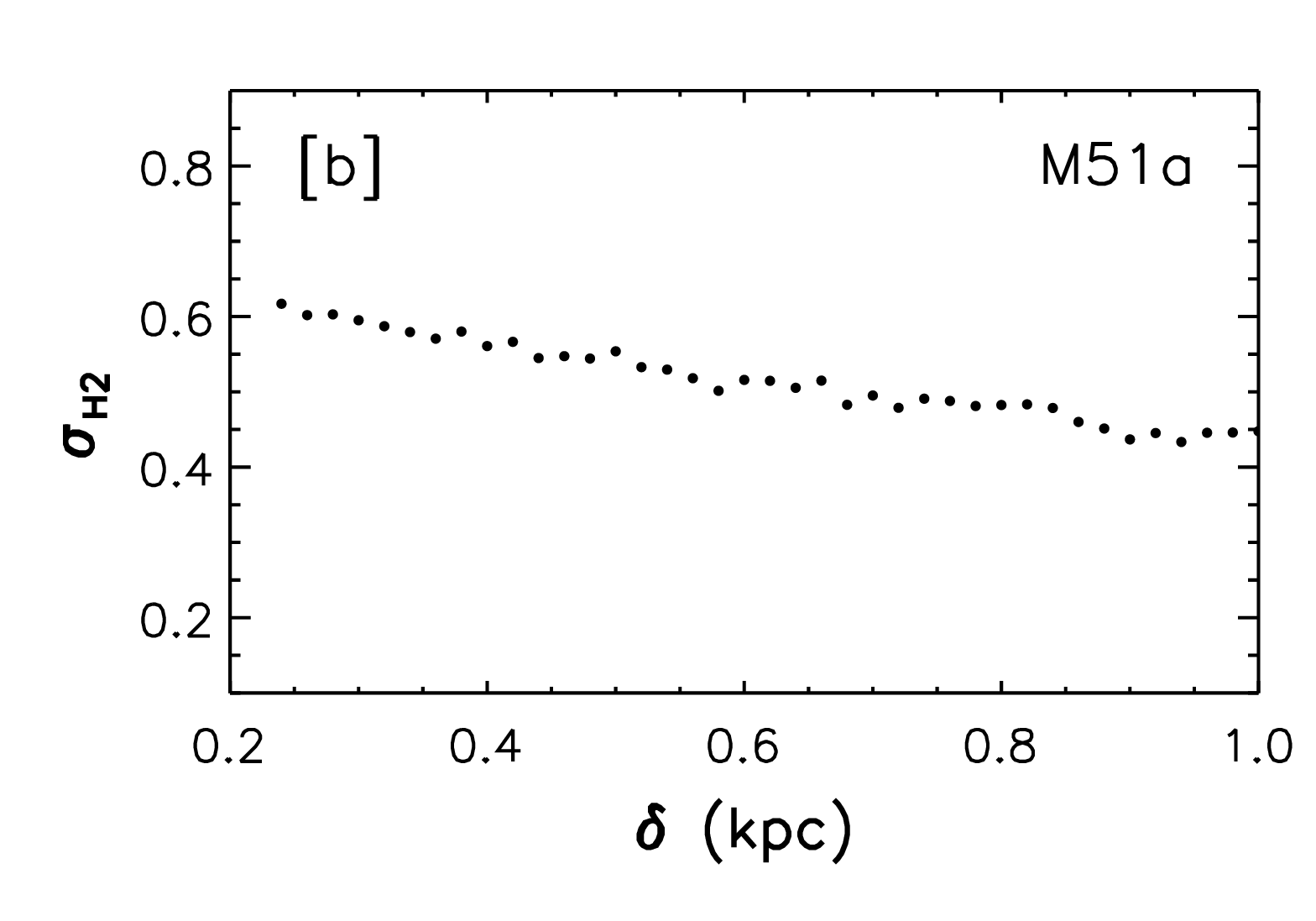}%
    \includegraphics[scale=.4,clip,trim=2.5cm 0cm 0cm 0cm]{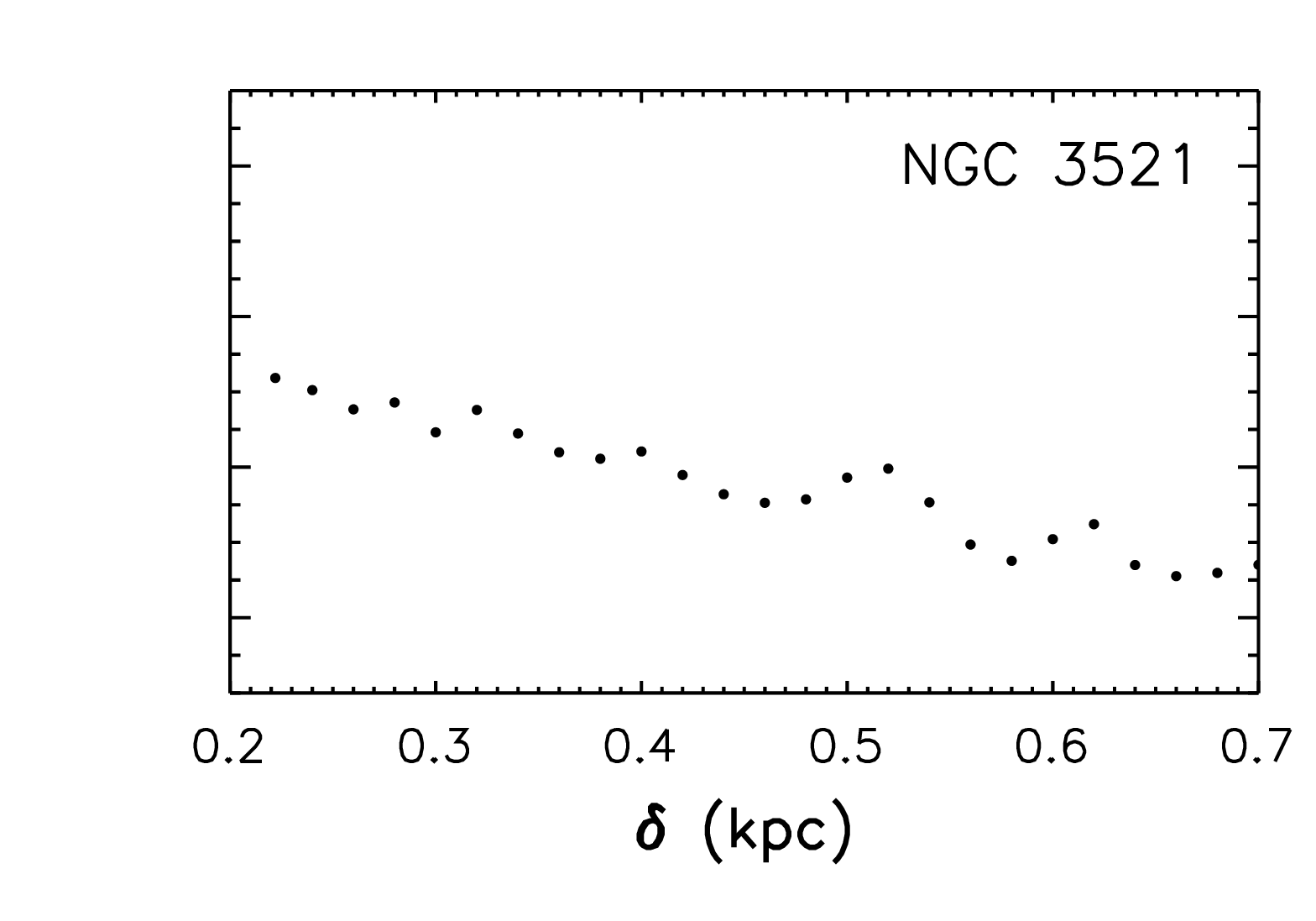}
    \includegraphics[scale=.4,clip,trim=0cm 0cm 0cm 0cm]{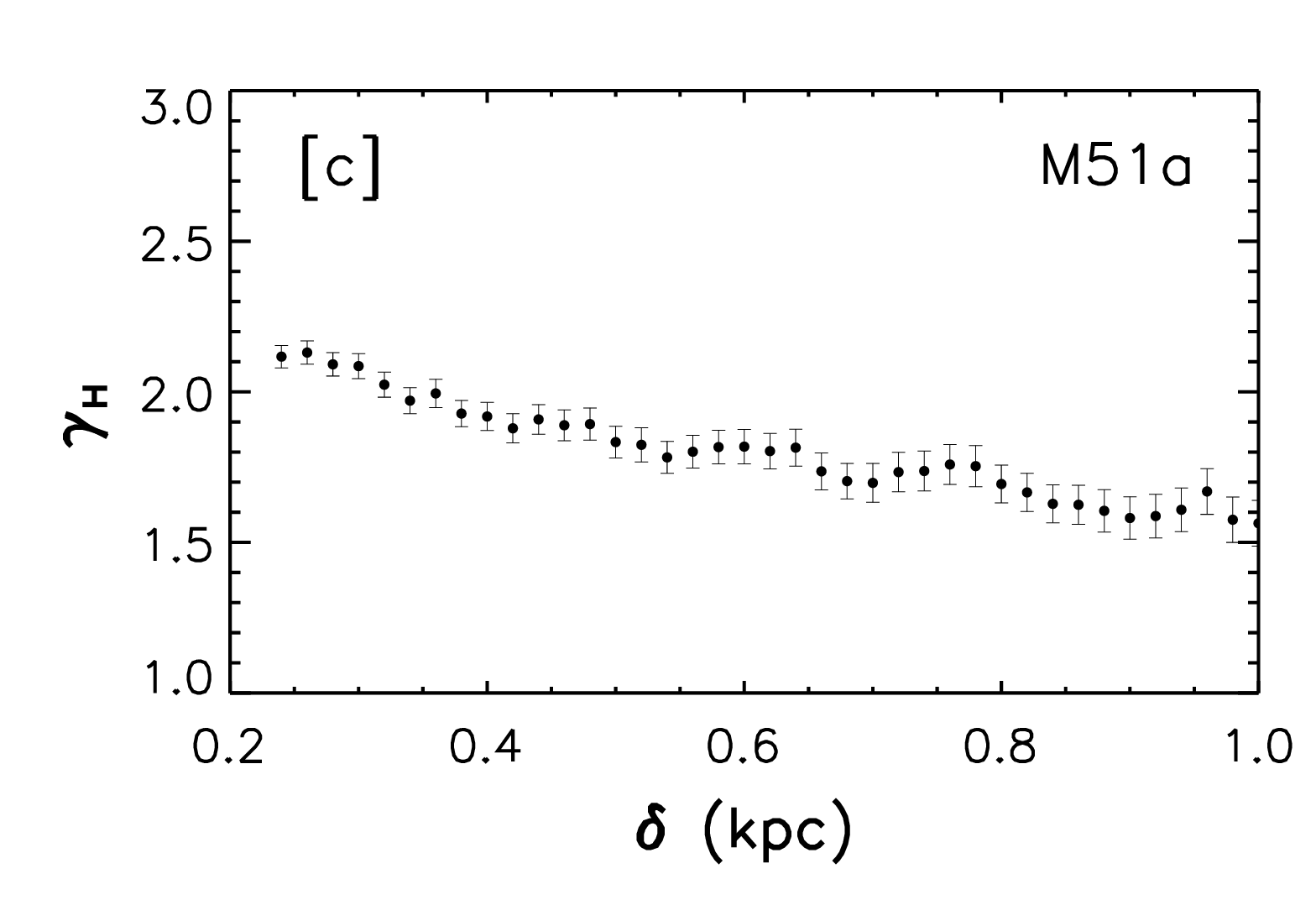}%
    \includegraphics[scale=.4,clip,trim=2.5cm 0cm 0cm 0cm]{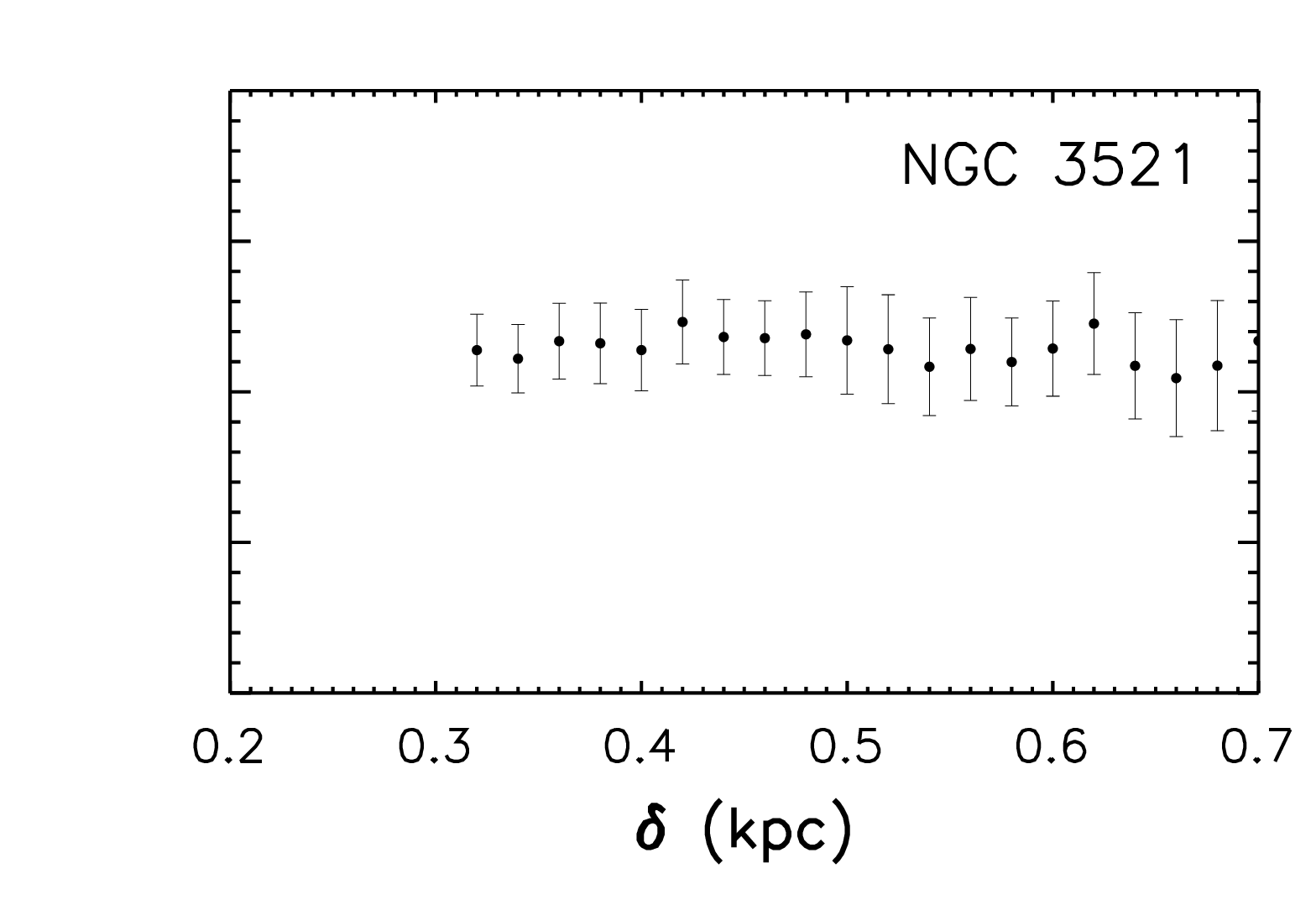}
    \includegraphics[scale=.4,clip,trim=0cm 0cm 0cm 0cm]{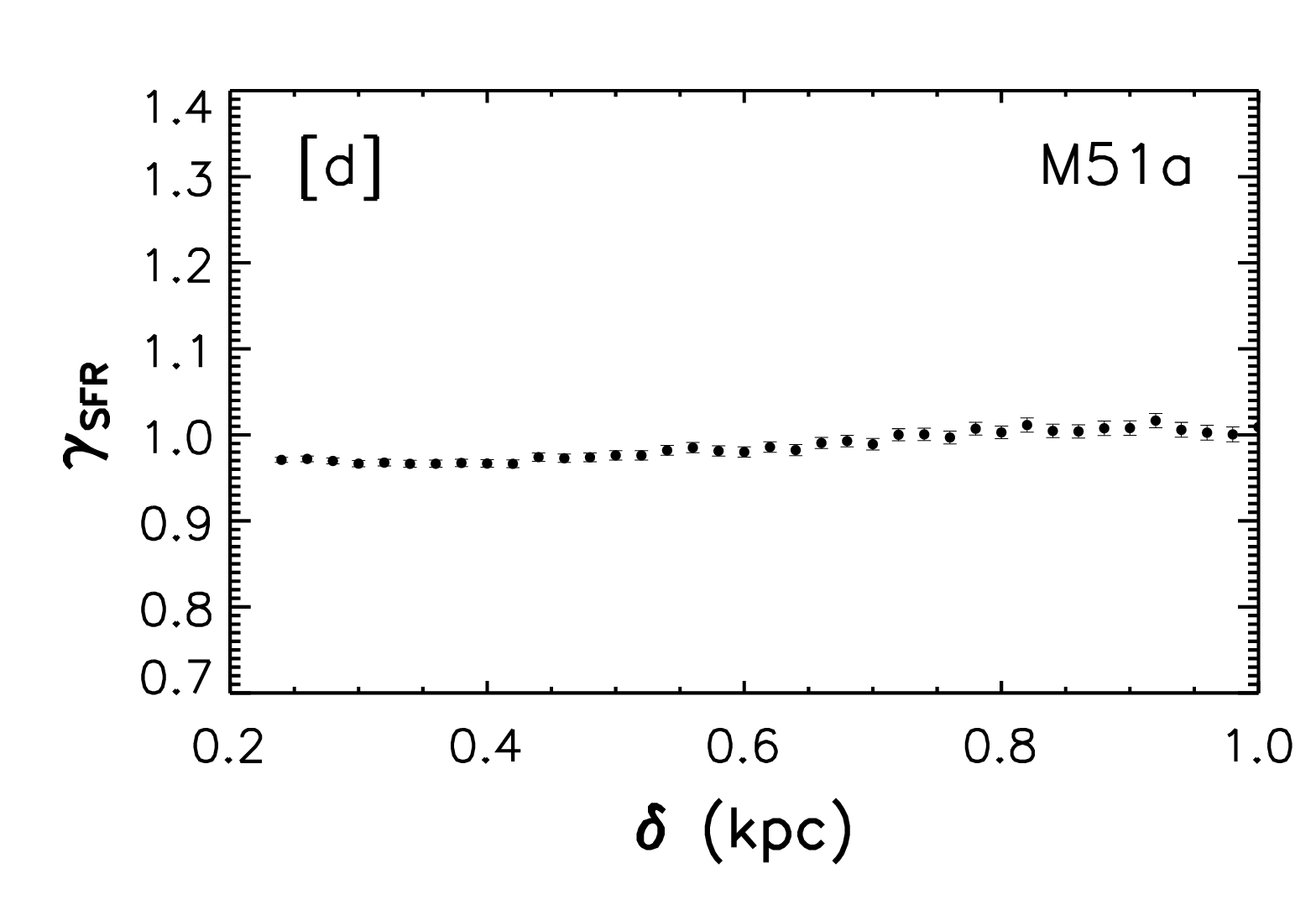}%
    \includegraphics[scale=.4,clip,trim=2.5cm 0cm 0cm 0cm]{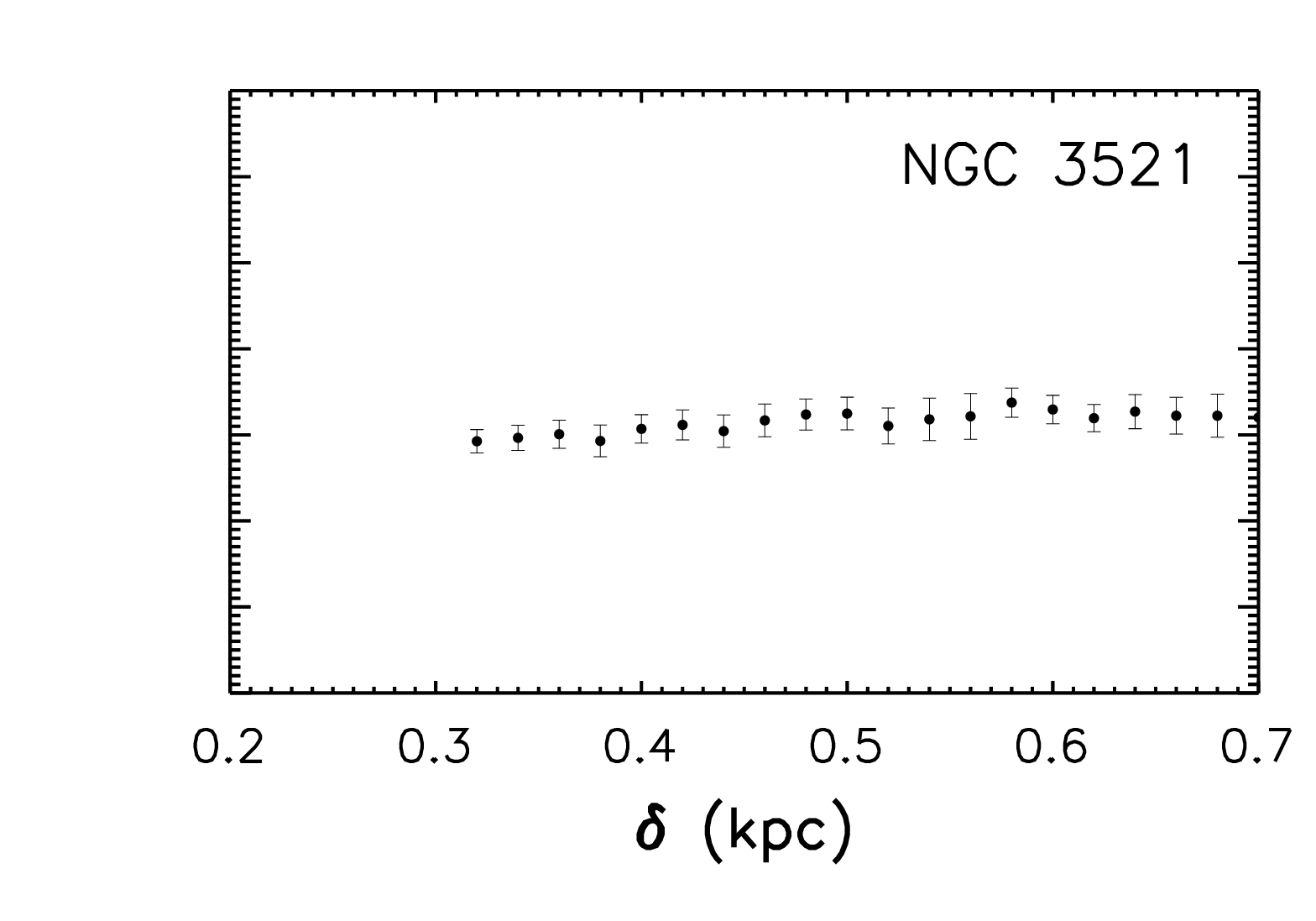}
    \caption{Best-fit parameters as a function of apparent resolution ($\delta$): 
[a] the power index of molecular-only S-K law; [b] the r.m.s. dispersion of the data
about the best-fit molecular S-K law in unit of dex; [c] the power index of the total-hydrogen 
S-K law (requiring 3$\sigma$ or better detection in both H {\sc i} and CO maps); 
[d] the power index of FUV+24 $\mu$m SFR vs. H$\alpha$+24 $\mu$m SFR correlation. 
Note that from [a] to [c] the SFR is derived from H$\alpha$+24 $\mu$m luminosity.}
\label{fig:res_dep}
\end{figure*}

We now investigate the scaling relationships at different spatial resolutions. The dependence of best-fit 
parameters on spatial resolution is shown in Figure~\ref{fig:res_dep}, where a series of power-law indices 
for M51a and NGC 3521 are shown in parallel. The range of spatial scales is different for 
these two objects due to their difference in inclination which yields a factor 2.5 smaller apparent minor
axis in NGC 3521 ($\sim$3 kpc across) than in M51a.
Investigations on scales larger than a quarter of this length will be of little significance because of 
the paucity of data points and of including a too large fraction of the galaxy. Moreover, the high inclination 
of NGC 3521 makes the removal of the background non-star-forming light more complicated, which can affect our results
especially on relatively larger scales. Considering the inclination of NGC 3521, 700 pc actually corresponds
to a physical scale of $\sim$2 kpc, which we take as a convenient limit for our study. The best-fit parameters 
of the molecular S-K law in M51a and NGC 3521 at different resolutions are reported in Table~\ref{tab:results_m51} 
and Table~\ref{tab:results_n3521}, where SFR is derived from the weighted sum of H$\alpha$ and 24 $\mu$m
luminosities.

Figure~\ref{fig:res_dep} demonstrates that $\gamma_{\rm SFR}$ has little dependence on spatial resolution
in both galaxies, varying between 0.97 and 1.01 in M51a and between 0.99 and 1.04 in NGC 3521. The weak 
ascending trend of $\gamma_{\rm SFR}$ is caused by the larger scatter at the faint end as discussed in the 
methodology section, and the uncertainty in background removal 
may also play a role. The FUV+24 $\mu$m luminosity is therefore also a good SFR tracer as long as the diffuse 
emission component is properly removed.

Figure~\ref{fig:res_dep} also reveals that the power-law indices of both the molecular and the total hydrogen 
S-K law decrease monotonically as the spatial scale increases. Specifically, as the resolution decreases from 
230 pc to 1 kpc, $\gamma_{\rm H_2}$ decreases from 1.9 to 1.2, implying higher star formation efficiency (defined as 
the SFR-to-gas surface density ratio) in the denser regime of molecular gas on all sub-kpc but super-GMC scales. 
For the total gas, we find $\gamma_{\rm H}$ 
decreases from 2.1 to 1.6 on the whole range of scales. At $\sim$500 pc resolution, we find steeper slopes
($\gamma_{\rm H_2}\sim1.5$, $\gamma_{\rm H}\sim1.8$) than K07 ($\gamma_{\rm H_2,K07}\sim1.4$, $\gamma_{\rm H,K07}\sim1.6$),
mainly because the new CARMA + NRO45 data is more sensitive than the BIMA SONG map by a factor of $\sim$2,
and has collected more diffuse emission from CO, pushing the low density end to higher values. 
A even steeper slope would be found if we followed B08, assumed no molecular gas outside the 
star-forming disk and fit the scaling relation on the whole isophotal area defined by $D_{25}$. 

In contrast to M51a, the high inclination of NGC 3521 results in blending of star formation 
sites which complicates the removal of diffuse stellar/dust emission, and can affect our result 
on relatively larger scales. Figure~\ref{fig:res_dep} shows that the slopes of the S-K law of both 
the total and molecular gas and of the SFR indicator comparison for NGC 3521 are less certain and stable than 
M51a, indicating higher uncertainties in the local background subtraction. In addition, the FoV of our CARMA 
observation covers only about half of the $D_{25}$ scale, missing the information from the outskirt of the 
disk where the transition from dominant molecular phase to dominant atomic phase in the cold gas occurs. 
This shortcoming tends to reduce the amount of data points in the regime slightly denser than 
$\rm \Sigma_{H}\sim10~M_{\odot}~yr^{-1}~kpc^{-2}$ and could have steepened the resulting slope. Bearing in 
mind these potential issues, we still observe a descending trend similar to that of M51a in both $\gamma_{\rm H_2}$ and 
$\sigma_{\rm H_2}$. However, both quantities are smaller than in M51a at a given resolution: $\gamma_{\rm H_2}$ is 
flatter by $\sim0.3$, and $\sigma_{\rm H_2}$ smaller by $\sim0.15$. The power-law index of $\gamma_{\rm H}$ in 
NGC 3521 fluctuates weakly within the errors and persists between 2.1 and 2.2, implying very weak or no 
dependence on the resolution. However, as we have mentioned above, this may due to the absence of the CO data 
in the outer star-forming disk which hinders a complete sampling over the range of gas surface density.

The scatter or dispersion of the S-K law, quantitatively defined as the r.m.s. of the fitting residuals 
measured perpendicular to the best-fit line, is an important parameter indicating the validity of the S-K law.
The study of K07 suggests that the 500-pc-scale total-gas S-K law has a scatter almost twice as large of 
the 2-kpc-scale measurements in M51a, which becomes the first hint of the existence of some scale below which 
the total-gas S-K law breaks down. Figure~\ref{fig:res_dep}-b reveals the monotonic decreasing
trend of the scatter ($\sigma_{\rm H_2}$) about the best-fit molecular S-K law as the resolution becomes 
lower on sub-kpc scales in both our program galaxies. \citet{Momose10} analysed the CARMA + NRO45 CO data of 
NGC 4303 and find that the K-S law already breaks down at $\sim$250 pc resolution, and does not recover
unless the data are smoothed to a scale of 500 pc or larger, especially in the disk region of the galaxy. 
Converselly, \citet{Verley10} show that a loosely correlated total S-K law exists down to 180 pc in M33 (0.84 Mpc), and 
\citet{Onodera10} conclude that the S-K law of H$_2$ in M33 becomes invalid only at the scale of GMCs ($\sim$80 pc). 
Limited by the angular resolution, we are unable to reach scales below $\sim$200 pc, but significant scatter 
($\sigma_{\rm H_2}$=0.62 dex for M51a and 0.52 dex for NGC 3521) about the best-fit molecular S-K law is 
observed in both galaxies at their highest available resolution. When the linear size increases by a factor 
of 3 ($\sim$700 pc), the scatter decreases to 0.48 (M51a) and 0.26 dex (NGC 3521). Assuming the trend in
Figure~\ref{fig:res_dep} [b] extends to smaller scales, a rough extrapolation gives a scatter of close to 0.7 dex 
at 80 pc scale, enough to mask any trend. The smoothly increasing scatter suggests the absence of 
a {\it characteristic} breakdown scale of the S-K law, but the scaling relationship will be overcome by the 
large scatter gradually. Higher resolution datasets are required to definitively pin down this issue. 

As an exploratory experiment, we also attempt to find a common formulation that relates the observational 
results in the two galaxies, so that some hints for the underlying physical universality may be obtained. 
Since the molecular S-K law is what should be focused on on sub-galactic scales, $\gamma_{\rm H_2}$ is of 
interest in the first place. Furthermore, because the uncertainty in the photometric measurements is
typically at the level of $\sim$10--20\%, the dispersion about the best-fit S-K law is mainly intrinsic, 
as has been pointed out by previous studies \citep[K07;][]{Blanc09}, and $\sigma_{\rm H_2}$ is a 
quantity of physical importance. 

The uncertainties of the slopes $\gamma_{\rm H_2}$ in Tables~\ref{tab:results_m51} and \ref{tab:results_n3521}
are, in fact, fitting errors, and are consistent in value with those already reported in the literature
\citep[K07;][]{Blanc09}. These uncertainties are not fully related to the dispersions $\sigma_{\rm H_2}$,
since, for instance, a larger dynamical range on the data could decrease the slope uncertainty, even at 
constant $\sigma_{\rm H_2}$. Hence the relevance of the intrinsic data scatter $\sigma_{\rm H_2}$, which
we report as separate columns in Tables~\ref{tab:results_m51} and \ref{tab:results_n3521}.

\begin{figure*}
    \includegraphics[totalheight=2.5cm,angle=0,origin=c,scale=3.4]{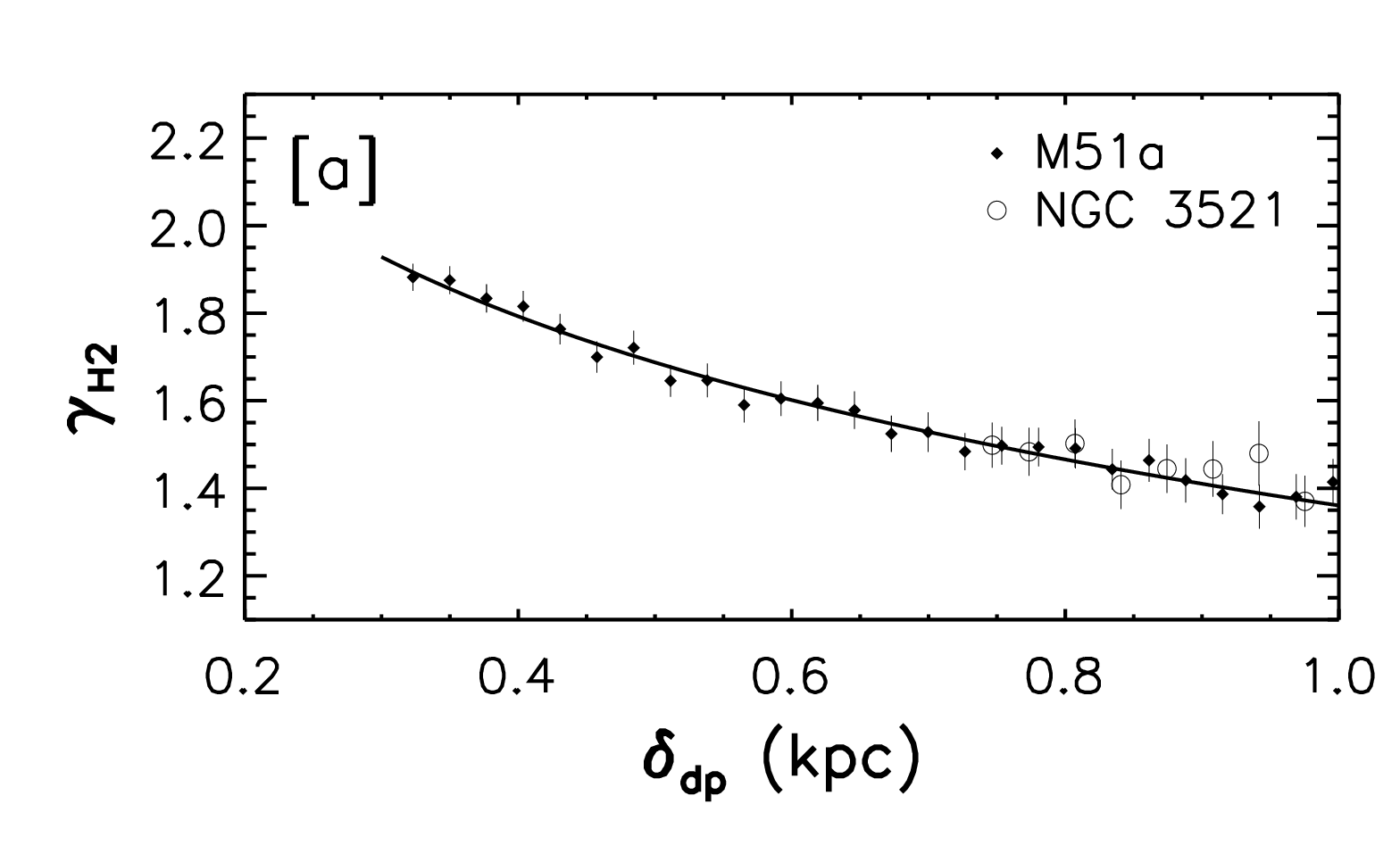}
    \includegraphics[totalheight=2.5cm,angle=0,origin=c,scale=3.4]{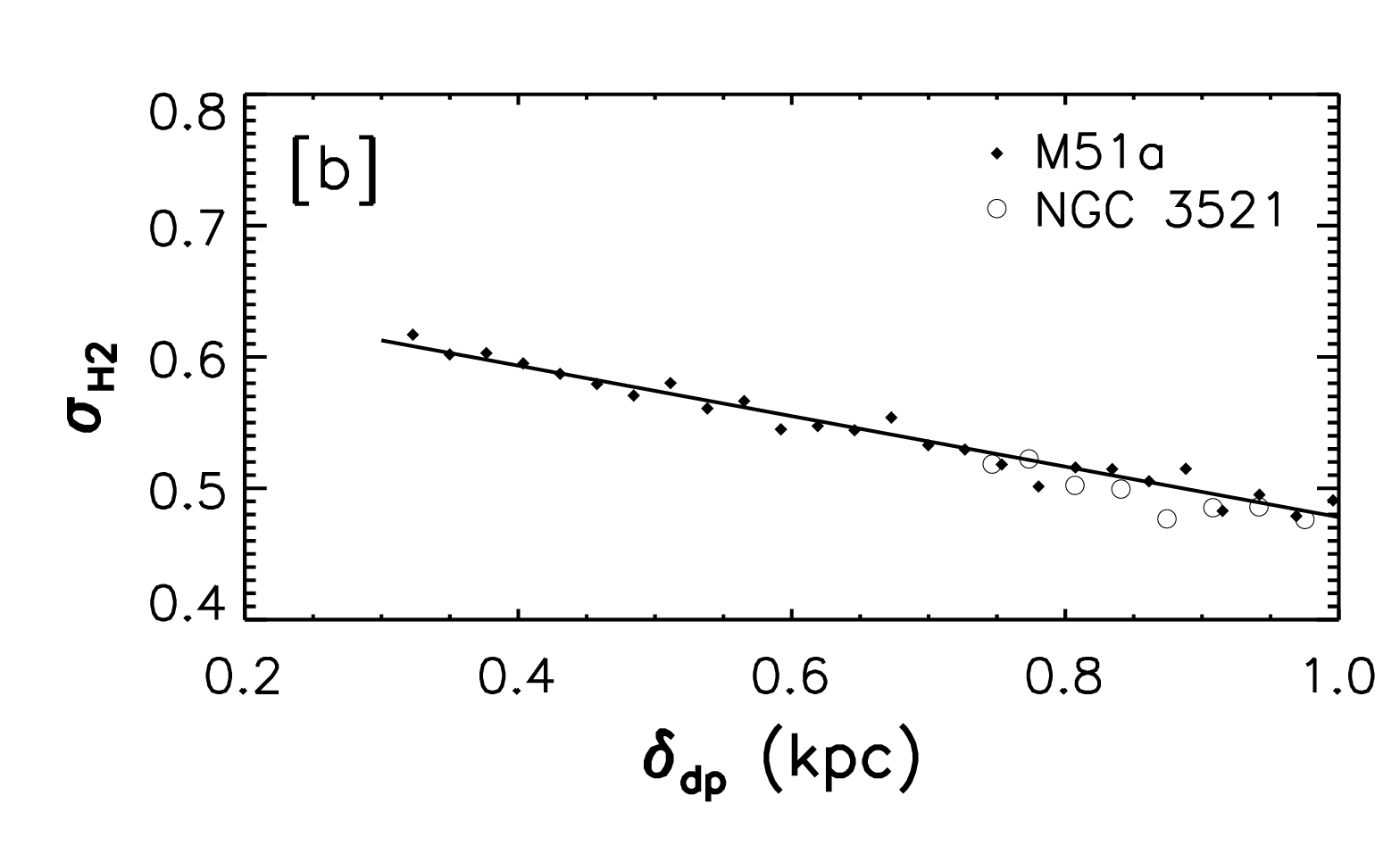}
    \caption{The power index ($\gamma_{\rm H_2}$) and r.m.s. dispersion ($\sigma_{\rm H_2}$) 
of the best-fit molecular S-K law in M51a and NGC 3521, plotted as a function of de-projected 
resolutions ($\delta_{\rm dp}$). Both quantities match remarkably well at a given $\delta_{\rm dp}$,
suggesting a shared behavior of sub-kpc molecular S-K law in them. Fitting to the data points
of M51a only, we find the logarithmic function $\gamma_{\rm H_2}=-1.09~\log~[\delta_{\rm dp}/{\rm kpc}]+1.36$ 
and the linear relation $\sigma_{\rm H_2}=-0.19~[\delta_{\rm dp}/{\rm kpc}]+0.67$ are able to depict the 
behavior of these two quantities in both galaxies remarkably well, as over-plotted with solid
lines in [a] and [b], respectively.}
\label{fig:co_rel}
\end{figure*}

As mentioned above, $\gamma_{\rm H_2}$ and $\sigma_{\rm H_2}$ are systematically smaller in NGC 3521 than 
in M51a. However, the resolution under consideration is the {\it projected} scale, and the comparison between 
the two galaxies should be made on the basis of {\it physical} or {\it de-projected} scales. We hence replot 
the dependence of $\gamma_{\rm H_2}$ and $\sigma_{\rm H_2}$ on $\delta_{\rm dp}$ ($\equiv \delta/\cos~i$) in
Figure~\ref{fig:co_rel}. As revealed by the two panels, on sub-kpc but super-GMC scales, both quantities are 
surprisingly consistent between the two galaxies. This remarkable result is a sign that we are approaching the 
scale where the physical origin of the S-K law is appearing. We are motivated to propose the following 
relations which hold in both galaxies:
\begin{equation}
\gamma_{\rm H_2}=A~\log~[\delta_{\rm dp}/{\rm kpc}]+B,
\end{equation}
and
\begin{equation}
\sigma_{\rm H_2}=C~[\delta_{\rm dp}/{\rm kpc}]+D,
\end{equation}
where $\sigma_{\rm H_2}$ is in dex, and $A=-1.09\pm0.05$, $B=1.36\pm0.02$, $C=-0.19\pm0.01$ and $D=0.67\pm0.01$. 
These best-fit parameters are determined using the data of M51a only, which are superior than those of 
NGC 3521 in many respects, as explained above. The values change minimally if the latter is included 
in the fits because of the impressive overlap among the data (Figure~\ref{fig:co_rel}). 
This overlap seemingly indicates a universal sub-kpc molecular S-K law that persists amongst spiral galaxies 
whose actual form depends only on the physical scale being considered. An extrapolation of the above expression 
for $\gamma_{\rm H_2}$ shows that a slope of unity requires a scale as large as $\sim2$~kpc, roughly the typical 
spacing of adjacent spiral arms in disk galaxies similar to M51a and NGC 3521. 
This has been verified in NGC 3521 (Figure~\ref{fig:res_dep}-a), where $\gamma_{\rm H_2}$ is consistent with
unity within the error bars at the apparent scale $\delta\sim700$ pc (Table~\ref{tab:results_n3521}), which
corresponds to a de-projected scale $\delta_{\rm dp}\sim2$ kpc. 

Similarly, we find that the $\gamma_{\rm H}$ -- $\delta$ correlation in M51a can be fitted to
\begin{center}
$\gamma_{\rm H}=-0.88~\log~[\delta/{\rm kpc}]+1.60$.
\end{center}
For a projected scale $\delta$=1.80 kpc (corresponding to 1.85 kpc if $D$=8.2 Mpc is adopted as in K07), the 
extrapolation of the above function predicts $\gamma_{\rm H}$=1.36, well consistent with the slope $1.37\pm0.03$ 
found at this resolution by K07 for the total hydrogen S-K law using the 14-m Five College Radio Astronomical 
Observatory CO ($J$=1--0) map \citep{Lord90}. Despite this consistency, we are aware of the uncertainties in
deriving large scale information from the extrapolation of our sub-kpc study.

The discussion above leads to a conclusion that caution must be taken in interpreting observational results, 
especially those of galaxies with high inclinations, which makes the intended sub-kpc studies actually 
``super-kpc''. Linear or sub-linear molecular S-K laws may be found if spiral structures are not readily resolved.
The HERACLES CO ($J$=2--1) data that is used by B08 has a resolution of 11\arcsec~\citep{Leroy09}. 
This spatial resolution (430 pc at a distance of 8.03 Mpc where NGC 3521 is located, or 570 pc at 10.7 Mpc adopted 
by B08) translates to $\delta_{\rm dp}\sim1.5$ kpc, already close to the $\sim2$ kpc unity-slope scale, and is virtually 
incapable of resolving the spiral structure of NGC 3521, which is likely to be another important reason for 
the results in B08 in addition to preserving the local background in the SFR measurements. 

At our highest available physical resolution ($\delta_{\rm dp}$=320 pc) in M51a, the power indices of both the 
molecular and total hydrogen S-K laws are about 2 ($\gamma_{\rm H_2}=1.91\pm0.03$, $\gamma_{\rm H}=2.14\pm0.04$). 
A slope as steep has never been reached in spatially-resolved studies of extragalactic star-forming disks, but 
has been observed inside the Milky Way. For instance, \citet{Misiriotis06} derive both the gas/dust 
distribution and SFR from COBE/DIRBE data and find $\gamma_{\rm H}=2.18\pm0.20$; \citet{Gutermuth11} directly 
observed about 7000 young stellar objects (YSOs) and obtain a slope $\sim$2, where the gas density is traced by 
dust extinction. Our results in M51a are therefore very similar to these Galactic investigations, which seems 
to hint at an ``intrinsic'' S-K law with slope $\sim$2 in spiral galaxies, and the 
flatter power indices that have been found in other galaxies are actually a resolution effect. As a
minimum, we have added evidence that a super-linear relation is found in extragalactic investigations
supporting the recent results that the star formation efficiency positively correlates with the column density 
for Galactic molecular clouds \citep{Gutermuth11}. As a caveat, our study has been 
performed only on two galaxies, and more data with better sensitivity, resolution and broader FoV for a
larger sample of galaxies are required to confirm the results presented in this section.

\section{DISCUSSION}

\subsection{The Nature of the Diffuse Emission at all Wavelengths}

\begin{figure*}
    \includegraphics[scale=.4,clip,trim=0cm 0cm 0cm 0cm]{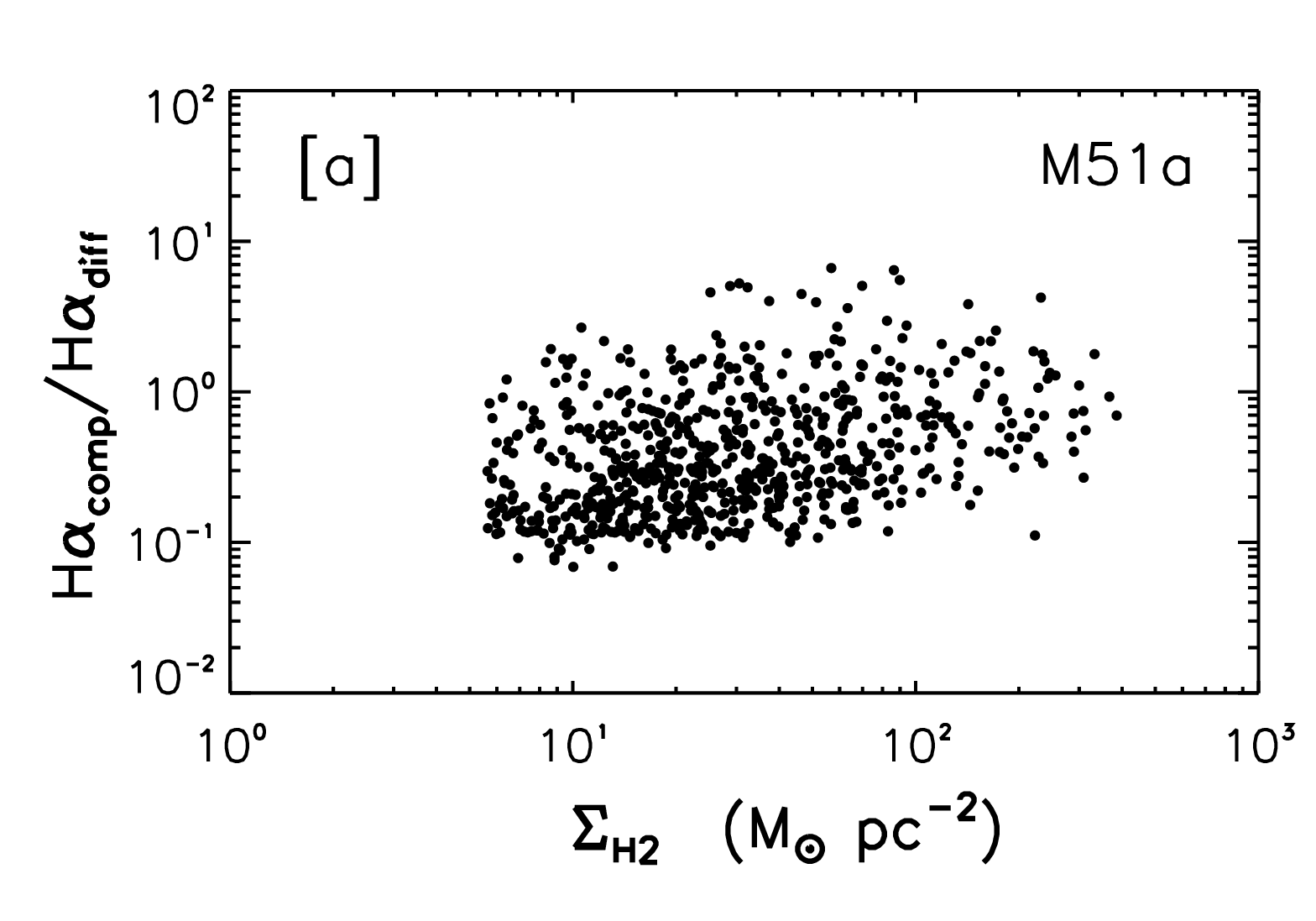}%
    \includegraphics[scale=.4,clip,trim=2.5cm 0cm 0cm 0cm]{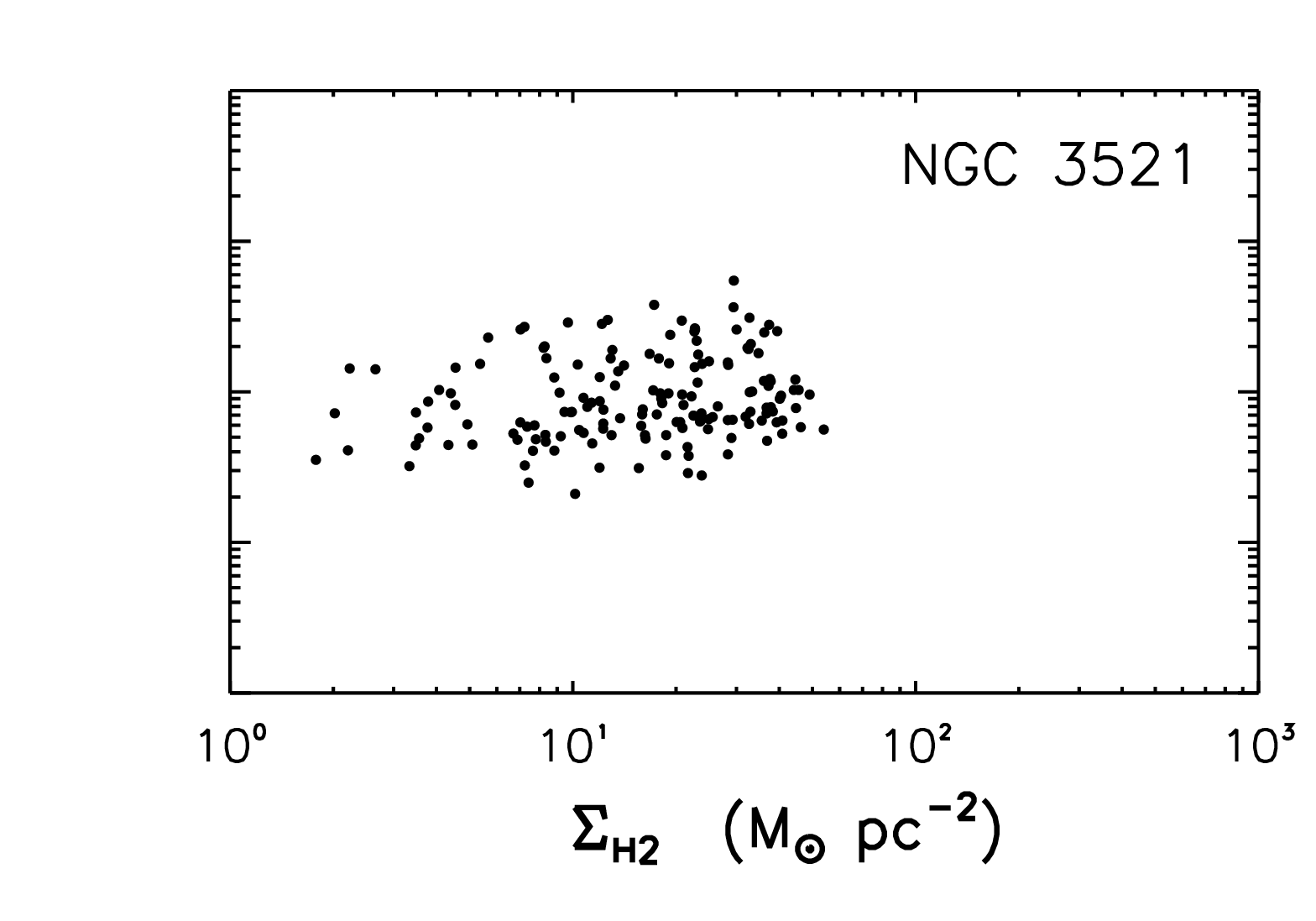}
    \hspace{4mm}
    \includegraphics[scale=.4,clip,trim=0cm 0cm 0cm 0cm]{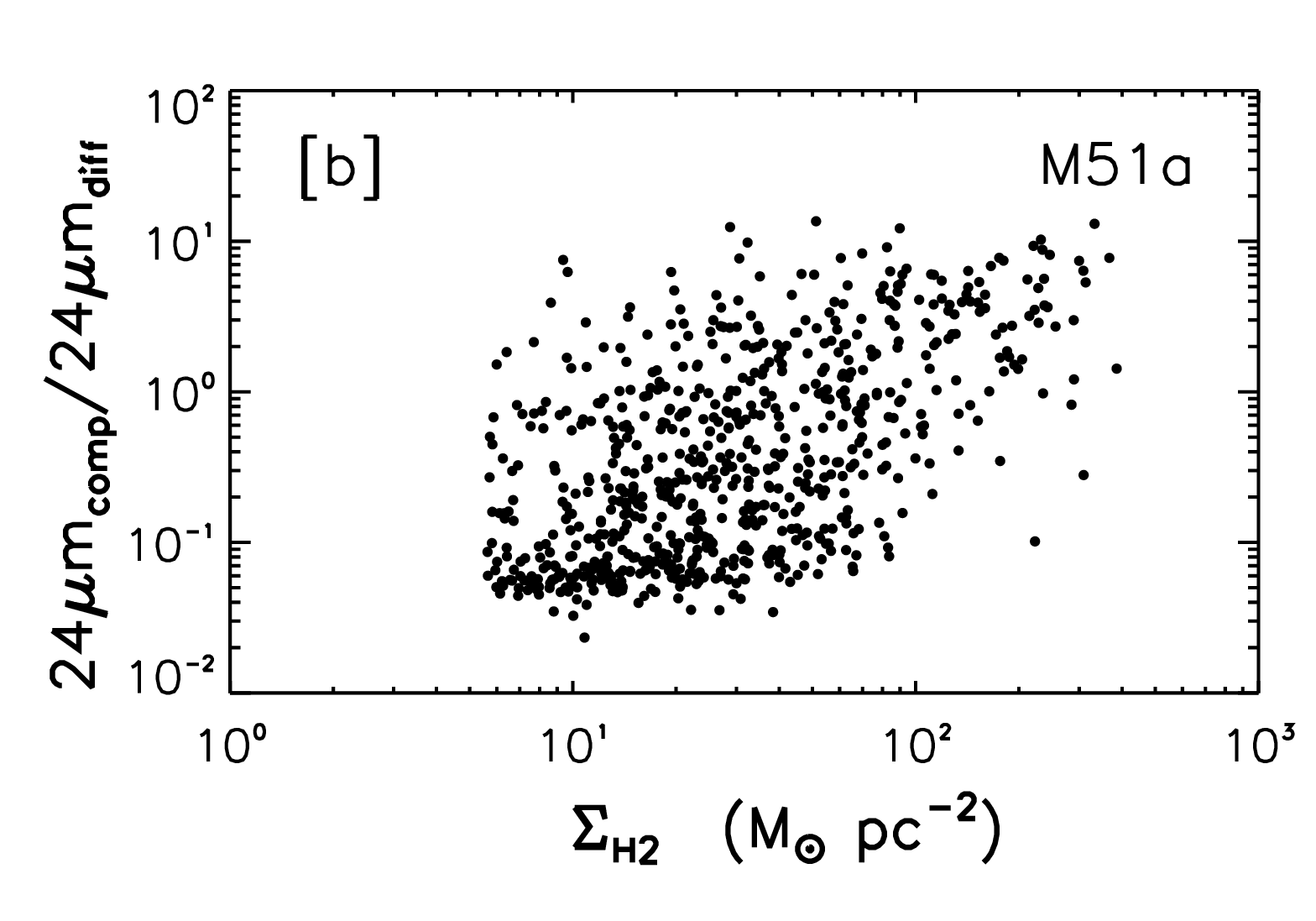}%
    \includegraphics[scale=.4,clip,trim=2.5cm 0cm 0cm 0cm]{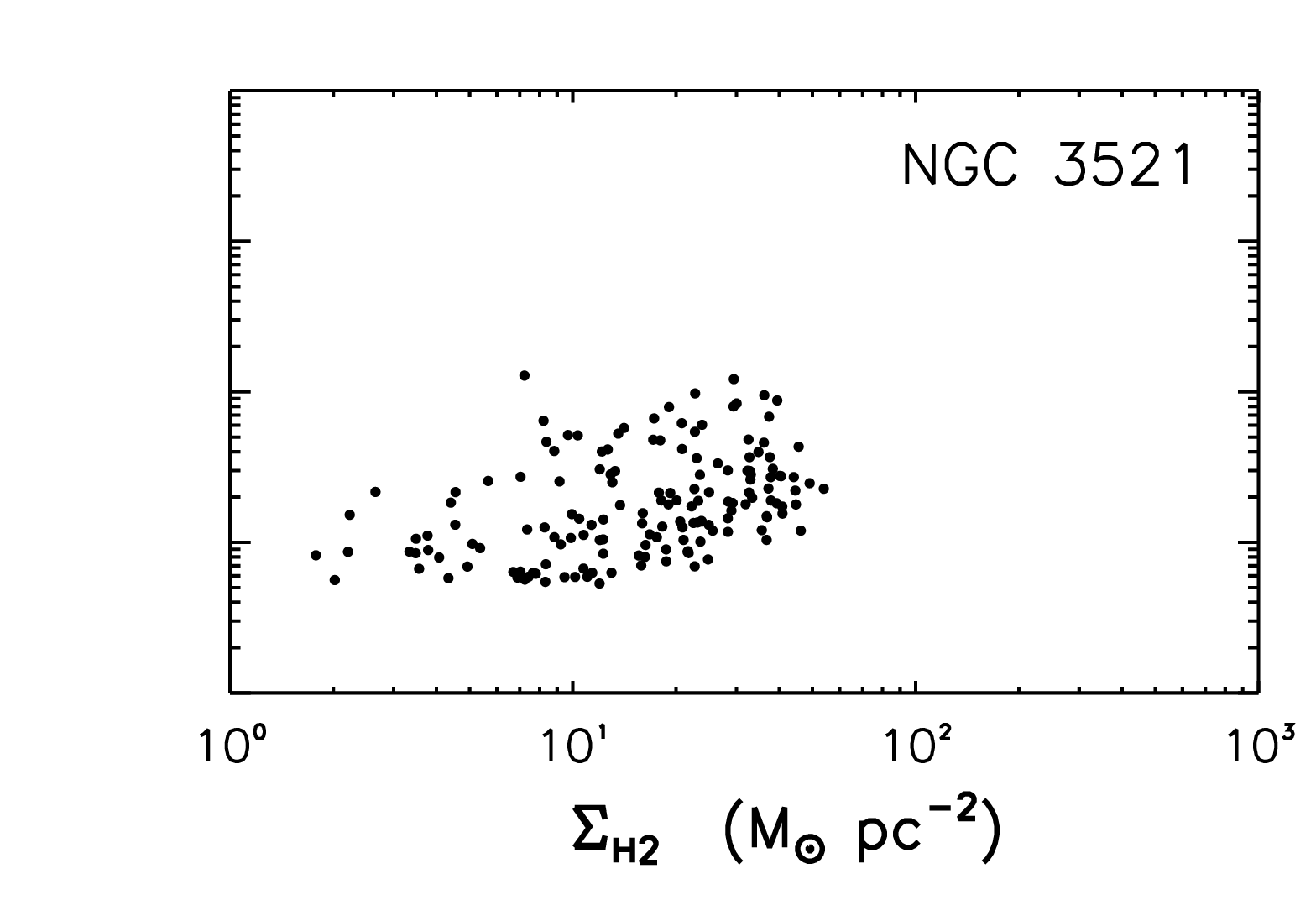}
    \hspace{4mm}
    \includegraphics[scale=.4,clip,trim=0cm 0cm 0cm 0cm]{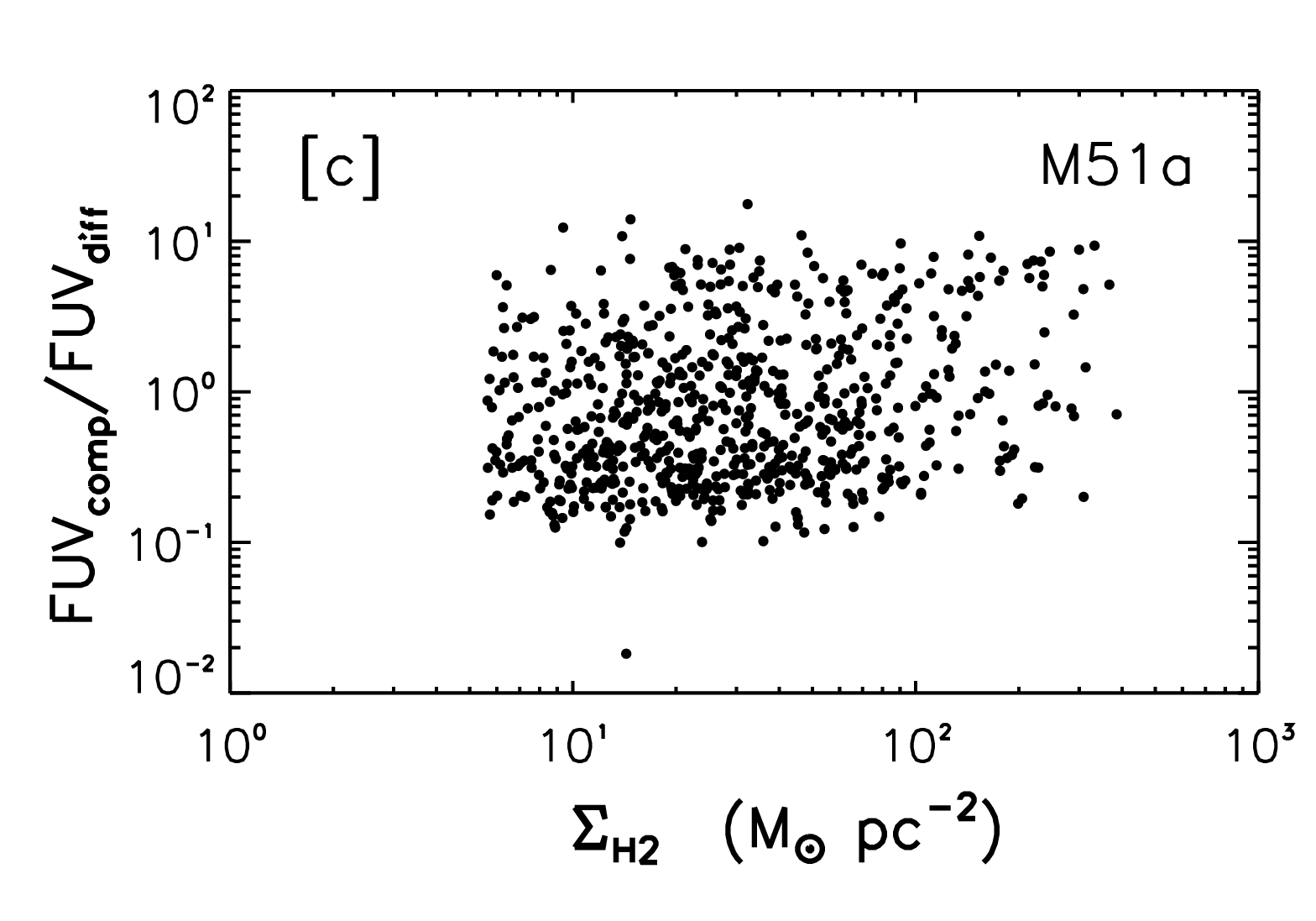}%
    \includegraphics[scale=.4,clip,trim=2.5cm 0cm 0cm 0cm]{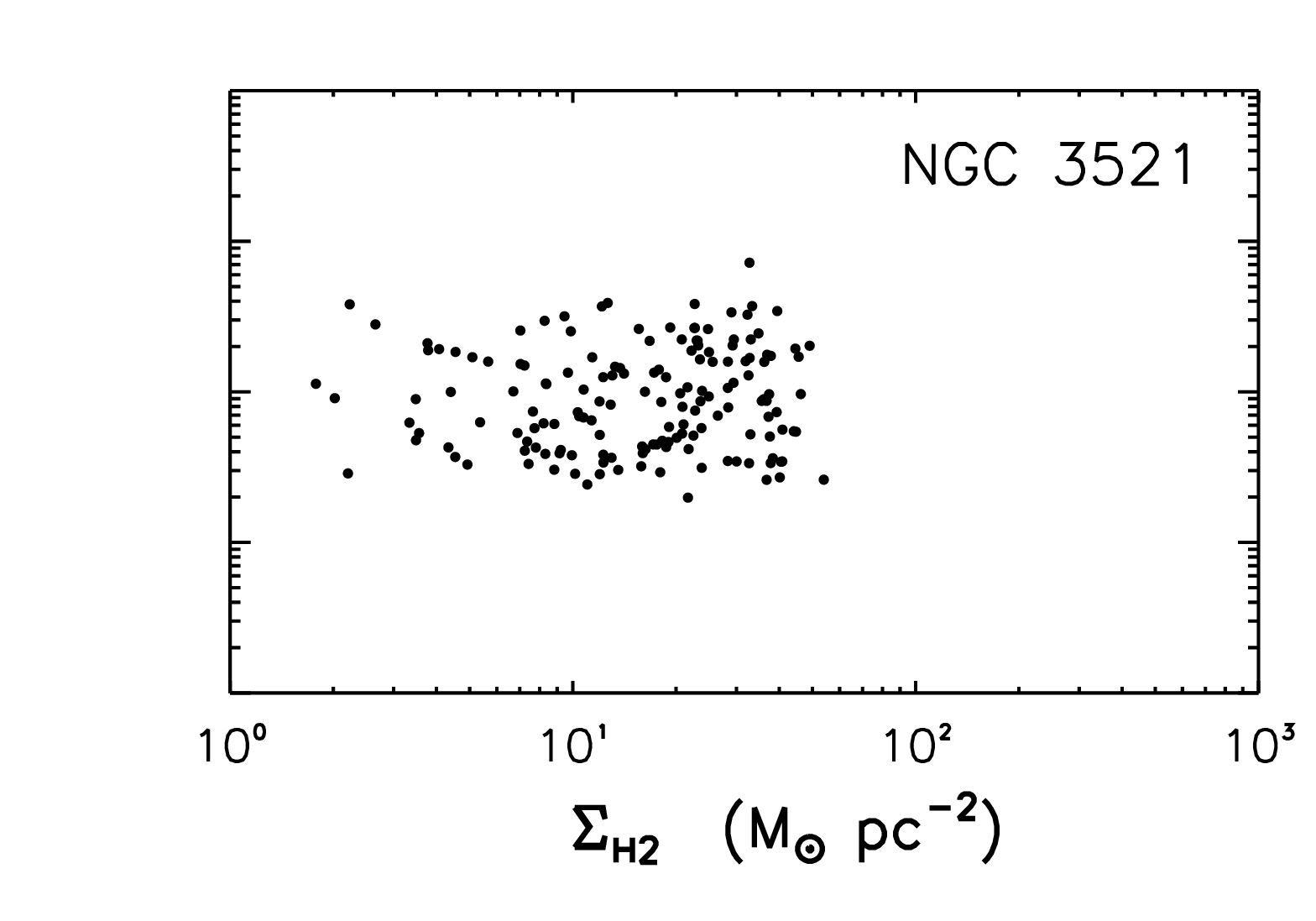}
    \caption{The compact-to-diffuse ratio of emission in: [a] H$\alpha$, [b] 24 $\mu$m 
and [c] FUV, ploted as a function of $\Sigma_{\rm H_2}$. Only $>$3$\sigma$ pixels are taken 
into account. For both M51a and NGC 3521, the projected resolution examined is 400 pc. Note 
the broader range of variation in 24 $\mu$m than in H$\alpha$ and FUV.}
\label{fig:diff}
\end{figure*}

Among the SFR images, the extended background contamination is the most significant at 24 $\mu$m. 
\citet{Popescu05} find that the IR-to-UV ratio is higher in the interarm regions of M101 than the
arm regions, which they interpret as due to dust heated by optical photons in the 
interstellar radiation field. In order to account for the observed SEDs of dust emission, a dust model 
must include a substantial population of ultrasmall grains or large molecules with the vibrational 
properties of PAH material and with sizes such that single-photon heating can excite the observed 
vibrational emission \citep{Draine07}.
Therefore, the mid-IR emission is produced not only by dust grains
heated by young stellar populations, but also, and in some areas
predominantly, by dust heated by older stars through single-photon
processes \citep[e.g.,][]{Draine07}. The latter component is
unassociated with current star formation, and is likely to dominate the
mid-IR emission in the faint, diffuse regions in galaxies.
As a quantification of the amount of diffuse light at each position in the two galaxies, we 
show in Figure~\ref{fig:diff} the ratio of the compact (background-subtracted) emission to the diffuse component 
(the subtracted background) in H$\alpha$, 24 $\mu$m and FUV as a function of H$_2$ mass surface density at 400 pc 
resolution (projected scale). It can be seen that, among the three wavelengths, the 24 $\mu$m emission shows 
the most marked trend for increasing compact-to-diffuse ratio with increasing $\Sigma_{\rm H_2}$ 
(especially in M51a). The variation of the compact-to-diffuse ratio is broader at 24 $\mu$m than H$\alpha$ and FUV
(especially in M51a), in line with the strongest influence from its diffuse emission. In general, the fact that 
young stars form in dusty galactic environment leads often to the prevalence of dust-obscured SFR (traced by IR) over 
unobscured SFR (traced by optical and UV), implying that the slope of the S-K law is mainly driven by the way 
the diffuse 24 $\mu$m emission is handled.

In the FUV data, the diffuse component is also considerably stronger than in the H$\alpha$ images 
\citep{Meurer95, Maoz96}, likely owing to the longer timescale 
over which UV continuum photons can be produced. As an example, it takes only 9 Myr for the H$\alpha$ emission 
of an instantaneous-burst population to diminish by two orders of magnitude, 
but takes about 100 Myr at 1500 \AA~where the FUV band of GALEX resides 
\citep{Leitherer99}. As a result of galaxy dynamics, over 100 Myr, the UV-emitting stars will have migrated 
away from their birth-site, and will have diffused across the galactic disk over an area at least 10 times 
larger than that of the ionizing-photon-emitting stars \citep{Chandar05, Pellerin07}. This results in a vastly 
reduced contrast between the arm and interarm regions of galaxies in FUV ($\sim$15:1 in M51a after extinction 
correction) relative to the H$\alpha$ ($>$30:1 in M51a), and also in a less obvious association between the 
location of the FUV emission and present-day star formation. Investigations on the properties of diffuse UV
emission have demonstrated that, the diffuse UV light is likely originating from evolved, disrupted stellar
clusters that has migrated away from their birth-sites \citep{Tremonti01, Chandar05}; any assumptions of in-situ 
star formation for the diffuse UV light implies a steep IMF which is either steeper than Kroupa at the high mass 
end ($\gtrsim$10-20 $M_{\odot}$) or truncated (down to 30-40 $M_{\odot}$) \citep[e.g.,][]{Tremonti01}. 

To estimate the impact of the diffuse UV light, we assume that
star formation proceeds at a roughly constant rate over 1 Gyr, and that
stars and molecular clouds remain closely associated over 
$\sim$30~Myr \citep{Blitz80,Elmegreen07}, in agreement with some observations
for the oldest stars found in GMCs \citep{Oliveira09}.
Stellar population synthesis models \citep[STARBURST99,][]{Leitherer99}
indicate that for constant star formation, 80\% of the GALEX FUV light
is produced by stars younger than 30 Myr, and 20\% by stars older than
30 Myr. If the older stars get dispersed over the galactic disk, their
mean dust attenuation will be lower than that of the younger stars, which
are clustered in the spiral arms. Using the opacity values of \citet{Holwerda05}, 
$A_{I,\rm arm}\sim$1.5~mag and $A_{I,\rm interarm}\sim$0.5~mag, and the 
conservative assumption of a mixed dust/star geometry, with a Milky Way 
extinction curve plus albedo \citep{Calzetti94, Calzetti00},
the contribution to the total {\em observed} FUV emission of a galaxy from 
stellar populations younger than 30~Myr is only about 60\%. Other
assumptions for the dust/star geometry and the effective attenuation will
lead to smaller fractions contributed by the young star populations to the
total observed FUV. In light of the above modeling, unless stellar populations 
significantly older than 30 Myr remain closely associated to their birth cloud, 
at least a 40\% fraction of ``diffuse'' UV light should be present.
Our strategy of diffuse emission subtraction presented here results in an
observationally determined diffuse FUV fraction of 44\% in M51a and 53\%
in NGC 3521. By virtue of its lower inclination, M51a has a more reliable
measurement, and our modeling and observation are therefore consistent.

Although removing the diffuse FUV emission is clearly a necessity for measuring 
SFRs, we have performed a test by making a SFR map of M51a with local background 
{\em preserved} for the FUV image but {\em subtracted} for 24 $\mu$m. Compared to our 
fiducial method of SFR tracing (H$\alpha$+24$\mu$m with the background removed in both bands),
this SFR map reduces the slope of the molecular S-K law $\gamma_{\rm H_2}$ from 1.42$\pm$0.05 
to 1.14$\pm$0.04 at 750 pc resolution, higher than the slope 0.83$\pm$0.04 obtained when both
the FUV and 24 $\mu$m backgrounds are preserved. Thus, keeping the FUV background recovers 
a slope roughly half-way between those obtained when either removing or preserving both 
backgrounds. However, we still recover non-linear trends at high spatial resolution:
$\gamma_{\rm H_2}$=1.39$\pm$0.02 at 250 pc resolution, when preserving the FUV background and 
removing it from the 24~$\mu$m image, to be compared with $\gamma_{\rm H_2}$=1.86$\pm$0.03, 
when the background is removed from both images.

The stronger diffuse emission in UV relative to the H$\alpha$ is the underlying reason for the deviation 
of the SFR$_{\rm FUV+24\mu m}$ vs. SFR$_{\rm H\alpha+24\mu m}$ correlation from a single power law at the 
faint end (see Figure~\ref{fig:m51_bima_comp}-c). As we have shown, this deviation can be calibrated by careful 
local background removal, whose necessity is thus justified by the need to include in our SFR indicators only 
the young stellar populations (and the dust they heat) involved in the {\it current} star formation activity. 
By including the local background contamination in their SFR estimates, B08 systematically overestimates the 
current SFRs at the faint end, which affects the physical interpretations of their high quality observations. 

The H$\alpha$ data are processed with our strategy as well, driven by the fact that a fraction of ionizing 
photons escape from H {\sc ii} regions and result in the diffuse emission from ambient ionized gas unrelated 
to the current star formation at a specific location \citep{Ferguson96, Ferguson98, Wang99, Heckman99}. Another 
reason is that pixel-by-pixel analysis is highly sensitive to the low level uncertainty in flat-fielding, 
continuum subtraction and sky background removal and requires careful handling. In fact, this technical 
difficulty can be circumvented by using integral field spectroscopy, as is done by \citet{Blanc09}.
The authors investigated the central 4$\times$4 kpc$^2$ of M51a at 170 pc resolution, with the diffuse H$\alpha$ 
emission corrected based on [S {\sc ii}]/H$\alpha$ line ratio, as an alternative to the local background 
removal in K07 and our work. Combining these data and the archival BIMA SONG CO map, they fit the data through 
a Monte Carlo approach with upper limits included and find a sub-linear slope $\gamma_{\rm H_2}=0.82\pm0.05$ for the 
molecular S-K law, similar to 0.84 obtained by B08. However, as they have pointed out, when an ordinary 
least-square (OLS) bisector linear fitting excluding upper limits (as throughout this paper, K07 and B08) is 
performed on their data, the slope steepens to 1.5, close to $\sim$1.4 found by K07. At our highest available 
resolution (230 pc) when the same BIMA SONG map is used, we find a similar power index $1.45\pm0.03$ if we only 
exclude all data points with significance below 1$\sigma$ as \citet{Blanc09} did, rather than our default 3$\sigma$.
Hence, they actually reach a conclusion consistent with K07 and this study, when noisy (below 1$\sigma$) CO
data are removed from their analysis.

\subsection{Robustness Tests}

In this work, we have performed our analysis in a customary manner, i.e. by working above a certain sensitivity 
threshold (3-$\sigma$). \citet{Blanc09} simulate the likely distribution of non-detections so that data below
1-$\sigma$ detection limits can be included in their power-law fitting. While this approach attempts to avoid 
data censoring, their strategy requires extremely accurate understanding and reliable modeling of the noise 
characteristics, which is a challenging enterprise especially for CO maps. 
Unfortunately, the current major limitation to spatially-resolved S-K law studies is the relatively low
sensitivity of the CO maps, and the real solution hinges on the higher quality of future CO data (through, 
e.g. ALMA). Our CARMA + NRO45 data for M51a is more sensitive than the BIMA SONG map that \citet{Blanc09} used by 
a factor of $\sim$2, and has higher image fidelity. Figure~\ref{fig:co_comp} compares these two maps at 230 pc 
scale (corresponding to the BIMA SONG resolution), and demonstrates that the CARMA + NRO45 observations recover 
slightly more flux at the faint end. Therefore, although different fitting techniques are at play, we have probed 
the CO emission to the same depth as \citet{Blanc09}, but at a much higher confidence level 
(2$\sigma$ vs. 1$\sigma$).

We now investigate the impact of imposing threshold values to our gas maps. As can be seen in 
Figures~\ref{fig:m51_bima_comp}-\ref{fig:n3521_sk}, the 3-$\sigma$ threshold has negligible impact on the SFR
maps. In fact, varying the SFR threshold from 2$\sigma$ to 4$\sigma$ changes the best-fit slopes minimally
(within 0.03). This remains true at all scales. Most of the data censoring occurs on the gas maps. 
We test the reliability of our fits by varying the CO map threshold from 1$\sigma$ to 5$\sigma$, and the results 
are tabulated in Table~\ref{tab:check_m51} and Table~\ref{tab:check_n3521}. As can be seen from the various
columns, decreasing the threshold of the CO map flattens the slope $\gamma_{\rm H_2}$ at all scales, suggesting
that the inclusion of noisy data has adverse effect on the determination of the S-K law. 
With future deeper CO maps, our current 1-$\sigma$ limits may become better detected, and enable a more accurate 
determination of the slope of the S-K Law at low CO masses. Using our current well detected data, with thresholds 
3-$\sigma$ or higher, the slope of the molecular S-K law is clearly super-linear.

Finally, one should be aware of the fact that because brightnesses at different wavelengths will often tend 
to correlate, causal relationships may be masked. In this work, this {\it richness effect} has 
been alleviated significantly when the underlying diffuse emission from dust and stars has been removed,
and is thus of much less importance than in the study of B08 (one may notice from Figure~\ref{fig:bkgd}
that the background-subtracted images no longer appear disk-like). The absence of a clear correlation 
between SFR and atomic hydrogen (especially when the local background component is subtracted, see
Figure~\ref{fig:m51_bima_comp}-d) is an implication that our analyses are not dominated by this effect.

Further progress on this problem of how to measure SFRs {\it locally} within galaxies will require a careful
modeling of the diffusion timescales of stellar populations in galaxies as they age, and the analysis of
both the direct stellar light (UV, H$\alpha$) and the dust-reprocessed light (IR) from these evolving populations.

\section{SUMMARY}

We have presented a case study of the sub-kpc Schmidt-Kennicutt law of star formation in two nearby galaxies, M51a 
(NGC 5194) and NGC 3521, using our new CARMA + NRO45 combined CO ($J$=1--0) maps of both galaxies with archival 
SINGS H$\alpha$, 24 $\mu$m, THINGS H {\sc i} and GALEX FUV data. Employing a new approach of local background 
determination, we use the same data sets as \citet{Kennicutt07} and \citet{Bigiel08} and quantitatively show that the 
contrasting results of the two studies on whether the molecular S-K law is super-linear or linear is a result 
of removing or preserving the local background, this being the sum of the diffuse emission in galaxy disks and 
bulges from stars and dust. We argue that in order to plausibly derive SFR maps of nearby galaxies, the local 
mid-IR and FUV background should be subtracted carefully, in order to isolate the currently star-forming regions.

Applying this strategy, we perform a pixel-by-pixel analysis on different physical scales and find the power 
index of molecular S-K law ($\gamma_{\rm H_2}$) to have similar behavior in these two galaxies. It is super-linear ($\gtrsim$1.5) at the highest available resolution ($\sim$220 pc), and decreases monotonically as the resolution 
becomes lower. We also find in both galaxies that the scatter of the molecular S-K law $\sigma_{\rm H_2}$
monotonically increases as the resolution becomes higher, indicating a trend for which the S-K law breaks down 
below some scale simply by reaching a scatter as large as the trend to be measured. Both $\gamma_{\rm H_2}$ and 
$\sigma_{\rm H_2}$ are systematically larger in M51a than in NGC 3521. 
However, when plotted against the {\it de-projected} scale, both quantities become impressively consistent between 
the two galaxies, tentatively suggesting that the measured molecular S-K law in spiral galaxies may depend only 
on the size of the physical scale considered, but does not vary from galaxy to galaxy.

\acknowledgments

The authors thank an anonymous referee for comments that have helped greatly improve this paper.
This work has been partially supported by the NASA ADP grant NNX 10AD08G.
We would like to thank the THINGS team and specifically Fabian Walter for providing their H {\sc i} maps of 
NGC 3521 and M51a. The Nobeyama 45-m telescope is operated by the Nobeyama Radio Observatory, a branch of the 
National Astronomical Observatory of Japan. Support for CARMA construction was derived from the Gordon and Betty 
Moore Foundation, the Kenneth T. and Eileen L. Norris Foundation, the James S. McDonnell Foundation, the Associates 
of the California Institute of Technology, the University of Chicago, the states of California, Illinois, and 
Maryland, and the National Science Foundation. Ongoing CARMA development and operations are supported by the 
National Science Foundation under a cooperative agreement, and by the CARMA partner universities.
\vspace{3 mm}

\clearpage

\clearpage

\begin{table}[ht]
\caption{\rm Best-fit parameters of the sub-kpc molecular S-K law in M51a (3-$\sigma$ thresholds are applied),
expressed as $\Sigma_{\rm SFR (H\alpha+24\mu m)} = A~\Sigma_{\rm H_2}^{\gamma_{\rm H_2}}$ with an r.m.s. 
dispersion $\sigma_{\rm H_2}$ about the best-fit power scaling.}
\centering
\begin{tabular}{c  c  c  c}
\hline\hline
Resolution (kpc)    &  $\gamma_{\rm H_2}$	&	$A$	&	$\sigma_{\rm H_2}$\\ 
\hline
0.25	&	1.86$\pm$0.03	&	$-5.49\pm$0.05	&	0.60	\\
0.30	&	1.82$\pm$0.03	&	$-5.36\pm$0.06	&	0.60	\\
0.40	&	1.65$\pm$0.04	&	$-4.97\pm$0.06	&	0.56	\\
0.50	&	1.52$\pm$0.04	&	$-4.71\pm$0.07	&	0.55	\\
0.60	&	1.49$\pm$0.05	&	$-4.62\pm$0.07	&	0.52	\\
0.70	&	1.36$\pm$0.05	&	$-4.36\pm$0.08	&	0.50	\\
0.80	&	1.32$\pm$0.05	&	$-4.24\pm$0.07	&	0.48	\\
0.90	&	1.26$\pm$0.05	&	$-4.14\pm$0.08	&	0.44	\\
1.00	&	1.21$\pm$0.06	&	$-4.08\pm$0.09	&	0.45	\\
[1ex]
\hline
\end{tabular}
\label{tab:results_m51}
\end{table}

\begin{table}[ht]
\caption{\rm Best-fit parameters of the sub-kpc molecular S-K law in NGC 3521. Symbols are the same as
for M51a.}
\centering
\begin{tabular}{c   c   c   c}
\hline\hline
Resolution (kpc)    &  $\gamma_{\rm H_2}$	&	$A$	&	$\sigma_{\rm H_2}$\\ 
\hline
0.25	&	1.41$\pm$0.06	&	$-4.48\pm$0.08	&	0.50	\\
0.30    &	1.40$\pm$0.06	&	$-4.43\pm$0.09	&	0.45	\\
0.40    &	1.35$\pm$0.09	&	$-4.32\pm$0.12	&	0.42	\\
0.50    &	1.33$\pm$0.12	&	$-4.25\pm$0.16	&	0.39	\\
0.60    &	1.19$\pm$0.11	&	$-4.05\pm$0.15	&	0.30	\\
0.70    &	1.24$\pm$0.16	&	$-4.12\pm$0.21	&	0.27	\\
[1ex]
\hline
\end{tabular}
\label{tab:results_n3521}
\end{table}

\begin{table}[ht]
\caption{\rm Best-fit slopes of the molecular S-K law when different $\Sigma_{\rm H_2}$ thresholds are 
applied but the SFR (traced by H$\alpha$+24 $\mu$m) threshold is fixed at 3-$\sigma$: the case of M51a.}
\centering
\begin{tabular}{c c c c c c}
\hline\hline
Resolution (kpc)    &   1-$\sigma$ & 2-$\sigma$ & 3-$\sigma$ & 4-$\sigma$ & 5-$\sigma$ \\ 
\hline
0.25	&	1.60	&	1.77	&	1.86	&	1.97	&	2.02	\\
0.30    &	1.51	&	1.67	&	1.82	&	1.88	&	1.97	\\
0.40    &	1.41	&	1.54	&	1.65	&	1.72	&	1.75	\\
0.50    &	1.33	&	1.44	&	1.52	&	1.60	&	1.66	\\
0.60    &	1.32	&	1.43	&	1.49	&	1.56	&	1.63	\\
0.70    &	1.20	&	1.27	&	1.36	&	1.46	&	1.48	\\
0.80    &	1.21	&	1.28	&	1.32	&	1.38	&	1.48	\\
0.90    &	1.10	&	1.18	&	1.26	&	1.29	&	1.32	\\
1.00    &	1.06	&	1.15	&	1.21	&	1.30	&	1.33	\\
[1ex]
\hline
\end{tabular}
\label{tab:check_m51}
\end{table}

\begin{table}[ht]
\caption{\rm Best-fit slopes of the molecular S-K law when different $\Sigma_{\rm H_2}$ thresholds are 
applied but the SFR (traced by H$\alpha$+24 $\mu$m) threshold is fixed at 3-$\sigma$: the case of NGC 3521.}
\centering
\begin{tabular}{c c c c c c}
\hline\hline
Resolution (kpc)    &   1-$\sigma$ & 2-$\sigma$ & 3-$\sigma$ & 4-$\sigma$ & 5-$\sigma$ \\ 
\hline
0.25	&	1.20	&	1.31	&	1.41	&	1.47	&	1.54	\\
0.30    &	1.19	&	1.34	&	1.40	&	1.44	&	1.49	\\
0.40	&	1.23	&	1.26	&	1.35	&	1.35	&	1.37	\\
0.50    &	1.22	&	1.22	&	1.33	&	1.33	&	1.33	\\
0.60    &	1.19	&	1.19	&	1.19	&	1.26	&	1.26	\\
0.70    &	1.24	&	1.24	&	1.24	&	1.24	&	1.35	\\
[1ex]
\hline
\end{tabular}
\label{tab:check_n3521}
\end{table}


\end{document}